\documentclass{aa}
\usepackage{etex}
\usepackage{graphicx}
\usepackage{listings}
\usepackage{amsmath,amssymb,mathrsfs}
\usepackage{empheq}
\usepackage{xcolor}
\usepackage{multirow} 
\usepackage{threeparttablex}
\usepackage{txfonts}
\usepackage{hyperref} 
\usepackage{rotating}
\usepackage{ulem}
\usepackage{natbib}
\usepackage{lineno}

\bibpunct{(}{)}{;}{a}{}{,}

\definecolor{dkgreen}{rgb}{0,0.6,0}
\definecolor{gray}{rgb}{0.5,0.5,0.5}
\definecolor{mauve}{rgb}{0.58,0,0.82}
\newcommand{\vr}{V_{R}}
\newcommand{\vsin}{V_{R,\sin}}
\newcommand{\vcos}{V_{R,\cos}}

\newcommand{\op}{\Omega_{p}}
\newcommand{\hr}{h_{R,1}}
\newcommand{\sigmamax}{\Sigma_{\max}(R_{\odot})}

\newcommand{\fullparam}{(\hr,\sigmamax,t, p)}

\newcommand{\kms}{\;\text{km\;s}^{-1}}
\newcommand{\kmskpc}{\;\text{km\;s}^{-1}\text{kpc}^{-1}}
\newcommand{\kpc}{\;\text{kpc}}
\newcommand{\mpc}{\;\text{M}_{\odot}\;\text{pc}^{-2}}
\newcommand{\gyr}{\;\text{Gyr}}
\newcommand{\fullunit}{\;(\text{kpc},\text{M}_{\odot}\;\text{pc}^{-2},\text{Gyr},\text{deg})}
\newcommand{\gaia}{\it Gaia}
\newcommand{\chired}{\chi^2_{\text{red}}}

\begin{document} 

   \title{Tracing the kinematic perturbations of the Milky Way spiral arms with APOGEE Data Release 17 and Gaia Data Release 3}
   
       \author{Xi-Can Tang \inst{1}
          \and Zhi Li \inst{1,2}\fnmsep\thanks{lizhi@shnu.edu.cn}
          \and Iulia T. Simion \inst{1}\thanks{isimion@shnu.edu.cn}
          \and Hao Tian \inst{3,4}
          \and Zhijian Luo \inst{1}
          \and Shuting Fan \inst{5}
          \and Zi-Qi Li \inst{1}
          }

   \institute{Shanghai Key Laboratory for Astrophysics, Shanghai Normal University, 100 Guilin Road, Shanghai 200234, China
        \and Shanghai Key Laboratory for Particle Physics and Cosmology, Shanghai, 200240, China
        \and Key Laboratory of Space Astronomy and Technology, National Astronomical Observatories, CAS, Beijing 100101, China
        \and Institute for Frontiers in Astronomy and Astrophysics, Beijing Normal University, Beijing 102206, China 
        \and Department of Astronomy, School of Physics and Astronomy, Shanghai Jiao Tong University 800 Dongchuan Road, Shanghai 200240, China
             }

   \date{Received February 5, 2026; accepted May 4, 2026}

\titlerunning{Tracing the kinematic perturbations of the Milky Way spiral arms}
\authorrunning{Xi-Can Tang et al.}

\abstract
{}
{We constrain the dynamical perturbations of the spiral arms in the Milky Way disk, based on the non-axisymmetric streaming motions of RGB stars 
revealed by APOGEE and $\gaia$.}
{We developed a revised steady-state radial-velocity response model that incorporates both the $\vsin$ and the dynamically important $\vcos$ components for a two-armed logarithmic spiral potential. The model was validated using orbit integrations with \texttt{AGAMA} and Bayesian parameter recovery with \texttt{dynesty}, and was applied to the smoothed two-dimensional radial-velocity field of RGB stars while accounting for Lindblad and corotation resonances.}
{The revised model reproduces the phase and amplitude of the mock radial-velocity field to the $\sim2\%$ level, substantially improving upon earlier $\vsin$-only formulations. Applied to the observational data, it yields a robust pitch angle of $p \simeq 10^\circ$ and a local surface density contrast of $\xi \simeq 5$--$18\%$ at the solar radius. The radial scale length is less well constrained ($\hr \simeq 40$--$50\kpc$) due to intrinsic parameter covariance. Resonance effects strongly shape the velocity field, affecting the fitting: the radial velocity becomes extremely large near the Lindblad resonances, whereas it vanishes close to the corotation resonance.}
{Our results demonstrate that including both the $\vsin$ and $\vcos$ terms is essential for a physically consistent interpretation of stellar streaming motions induced by a spiral potential. The observed kinematics constrain the spiral pattern speed to $\op \approx 10$--$20\kmskpc$.}


   \maketitle 
   \nolinenumbers

\section{Introduction} \label{sec:intro}
The morphology of the Milky Way spiral structure is still an open question \citep{churchell2009_rc, hou2014_spiral, xu2016_local, rezaei2018_rc_rg}. Studies have suggested both a two-arm \citep[e.g.,][]{xu2023_seg} and a four-arm \citep[e.g.,][]{reid2019_spiral,drimmel2024_p} grand-design spiral shape, complicated by several spurs and segments. The classification of the Local Arm, where the Sun resides, as a major arm remains contentious. Numerous studies have described it as a minor spur \citep{georgelin1976_spiral,SJT2020_bar_sa}, whereas \citet{xu2016_local, xu2018_spiral} suggested that its properties are comparable to those of major arms. On smaller spatial scales, the structure of the Local Arm is further complicated by coherent features in the local interstellar medium (ISM). The recently identified Radcliffe wave \citep[e.g.,][]{Alves2020_Radcliffe,li2022_wave,Konietzka2024_Radcliffe} is an osculating  kiloparsec-scale structure composed of molecular clouds and star-forming regions, indicating that the local segment of the arm departs from a simple planar spiral morphology.

The spiral structure can be mapped using various observational methods. In the solar neighborhood, young OB stars and associated molecular clouds are effective tracers due to their high brightness and their confinement to the spiral arm potential \citep{bobylev2013_ob, pantaleoni2021_ob}. At greater distances, classical Cepheids, and red giant branch (RGB) stars are commonly used owing to their reliable distance estimates, though care is needed in the outer disk to separate spiral features from the Galactic warp \citep{churchell2009_rc, rezaei2018_rc_rg, lemasle2022_cepheids}. In addition to these stellar tracers, recent advances in three-dimensional dust mapping using $\gaia$ mission have provided a new view of the Milky Way’s structure\citep{Green2019_dustmap,Lallement2022_dustmap,Zucker2023_dustmap,Zucker2025_dustmap,Edenhofer2024_dustmap,Barbillon2025_dustmap}. 
These studies reconstruct the distribution of interstellar dust and have significantly improved distance estimates, revealing coherent large-scale features associated with spiral arms. On the other hand, spiral structures can also be revealed through their dynamical imprints on stars and gas. Studies using perturbation theories \citep{eilers2020strength, khalil2024_vel} as well as numerical simulations \citep{Grand2012_nbody, Antoja2022_tidal,lz2022_gas,hunter_etal_24} have analyzed the non-axisymmetric velocity pattern due to the spiral potential, thereby constraining the spiral arm properties.

There are usually three major parameters to characterize a spiral pattern; namely, the pattern speed, $\Omega_p$, the pitch angle, $p$, and the surface density contrast between the arm and inter-arm regions. The pattern speed, $\Omega_p$, of spiral arms represents the angular velocity of the rotating spiral pattern. For the Milky Way, estimates of $\Omega_p$ vary widely, ranging from 17 to $30 \kmskpc$ 
 \citep{mishurov1997_op, dias2019_op, bobylev2023_op,Bobylev2025_op}. However, early density wave theory by \cite{lin1964_op} suggested a slower speed near $13 \kmskpc$, supported by later findings \citep{vallee2021_op, khalil2024_vel}. Using large-scale radial velocity data, \citet{eilers2020strength} proposed a constant pattern speed of $\Omega_p = 12 \kmskpc$, consistent with a two-armed spiral structure associated with the Local and Outer-Norma arms. Additionally, \citet{castro2021_op} analyzed open cluster distributions and reported a radial decline in pattern speed, ranging from approximately $50 \kmskpc$ in the inner Galaxy to $17 \kmskpc$ in the outer regions.

The pitch angle, $p$, characterizes the winding of spiral arms. A pitch angle of approximately $p\sim6^{\circ}$ is often adopted for two-armed models \citep{schmeja2014_p}, while four-armed models typically assume $p\sim13^{\circ}$ \citep{bobylev2014milky}. However, a single global pitch angle is unlikely to accurately describe all spiral arms. \cite{vallee2017_p} measured pitch angles for three inner-disk arms using two methods: the twin-tangent method yielded pitch angles of $12-15^{\circ}$ for the Sagittarius arm, $11-15^{\circ}$ for the Scutum arm, and $11-15^{\circ}$ for the Norma arm (including its Cygnus extension); the parallax method gave broader ranges of $6-19^{\circ}$, $6-20^{\circ}$, and $2-15^{\circ}$, respectively.

The surface density contrast between the arms and inter-arm regions can be directly revealed by star counts (e.g., \citealt{poggio2021_den}), although observational biases and sample selection effects need to be carefully considered. On the other hand, the radial and azimuthal stellar velocity patterns driven by a spiral perturbation in the disk can be constrained by dynamical models \citep{Antoja2011_kinematic}. Using the mean Galactocentric radial velocity,  \citet{siebert2012properties} applied a two-armed perturbation model and inferred a density enhancement with an amplitude of about 14$\%$ relative to the solar-neighbourhood background, while \citet{eilers2020strength} used a similar method to propose a smaller local surface density contrast of $10\%$ at the solar radius. Meanwhile, \citet{khalil2024_vel} found that the local contrast density depends on the number of spiral arms, with $9\%$ contrast for a three-arm spiral and $25\%$ for a two-arm structure.

The physical nature of the Milky Way's spiral structure remains actively debated. In the classical quasi-stationary density wave framework, spiral arms are modeled as long-lived, rigidly rotating global patterns \citep{lin1964_op,Roberts1969_density_theory,Bertin1996_densitywave,Shu_2016_ARAA}, and this approach has been widely adopted to interpret stellar and gas kinematics in the Milky Way \citep[e.g.,][]{siebert2012properties,vallee2021_op}. However, an alternative ``dynamic spiral'' scenario describes spiral arms as short-lived and recurrent features\citep{Dobbs2014_transient_theory, Sellwood2022_transient_theory}. In this picture, kinematic substructures revealed by $\gaia$, such as phase-space ridges \citep{Hunt2018_transient_theory}, disrupting and growing signatures \citep[][]{Baba_etal_2018,funakoshi_etal_2024}, as well as varying pattern speeds across different arms \citep{castro2021_op}, are better explained. In addition, tidally induced spirals driven by external perturbations can also generate nonequilibrium spiral features in the Galactic disk \citep{Laporte2019_tidal, Antoja2022_tidal}. A comprehensive review of this topic is given by \citet{Hunt2025_tidal}.

In this work, we revisit the kinematic imprints of spiral arms on the Galactic disk using APOGEE DR17 with {\tt StarHorse} distances \citep{queiroz2023starhorse} and $\gaia$ DR3 proper motions, together with a steady-state dynamical model inspired by \citet{eilers2020strength}. Our goal is to constrain the aforementioned three parameters by using RGB stars as luminous tracers. The structure of this paper is as follows. Section~\ref{sec:data} describes the RGB sample used in our analysis, as well as the construction of the observed mean radial-velocity map. Section~\ref{sec:model} introduces the steady-state toy model of the spiral arm perturbations. Section~\ref{sec:mock} investigates, based on orbit-integration results, how the physical parameters affect the model predictions and outlines the {\tt dynesty} fitting procedure. Sections~\ref{sec:result} and \ref{sec:dis} present the fitting results for RGB data and discuss them in the context of previous studies. Finally, Section~\ref{sec:summary} summarizes our main conclusions and provides prospects for future work. Detailed model expressions are provided in the Appendix.

\section{Data}\label{sec:data}

\begin{figure*}
    \centering
    \includegraphics[width=0.95\textwidth]{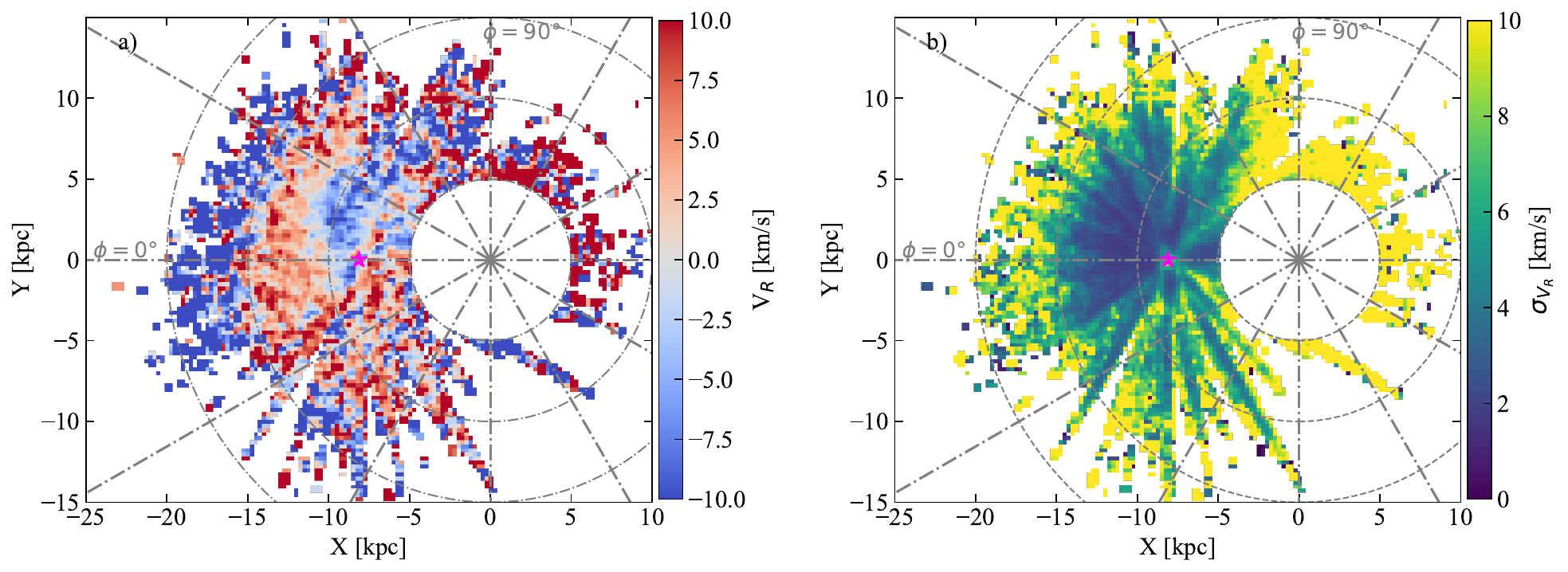}
    \caption{Radial velocity and corresponding uncertainty maps of RGB stars in APOGEE DR17 and Gaia DR3, smoothed over a spatial scale of 2.5 times the bin size. The magenta star marks the position of the Sun.}
\label{fig:xy_vr_sig}
\end{figure*}

To investigate the in-plane motions of disk stars and study the dynamical effects of the Milky Way's spiral arms, we selected a sample of RGB stars from APOGEE DR17 and $\gaia$ DR3 via the following query:
\begin{lstlisting}[language=sql]
SELECT *
FROM 
    "J/A+A/673/A155/apogee17" AS ap,
    "I/350/gaiaedr3" AS ga
WHERE 
    ap.GaiaEDR3 = ga.Source
    AND ap.logg50 > 0.0 AND ap.logg50 < 2.2
    AND ap.teff50 > 3000 AND ap.teff50 < 5500
    AND ap.dist50 / (ap.dist84 - ap.dist16) > 2.5
    AND ga.RUWE < 1.4;
\end{lstlisting}
Applying these criteria yields an initial sample of 125,082 RGB stars, which possess comprehensive and high-precision photometric and astrometric data, complemented by distance measurements from the StarHorse catalog \citep{Anders2020_starhorse,queiroz2023starhorse}.

Based on the Galactic longitude, $l$, and latitude, $b$, along with the distance, we mapped RGB stars into the three-dimensional Galactocentric Cartesian coordinate system ($X, Y, Z$). In this system, the positive $X$ axis points toward the Galactic center. The transformation is given by
\begin{equation}
\begin{pmatrix}
X\\Y\\Z
\end{pmatrix} 
= d 
\begin{pmatrix} 
\cos b\cos l\\ 
\cos b\sin l\\
\sin b
\end{pmatrix}
+\begin{pmatrix}
-R_\odot \\ 
0\\ 
Z_\odot    
\end{pmatrix},
\end{equation}
where $ R_\odot = 8.122 \pm 0.031 $ kpc \citep{abuter2018detection} and $ Z_\odot \approx 0.025 $ kpc \citep{juric2008milky} represent the solar distance from the Galactic center and its height above the Galactic midplane, respectively. To further analyze the kinematics of RGB stars, we transformed their coordinates from Cartesian coordinates ($X, Y, Z$) to left-handed Galactocentric cylindrical coordinates ($R,\phi$), where the radial distance and azimuthal angle are given by $R = \sqrt{X^2 + Y^2}$ and $\phi = \arctan\left(\frac{Y}{-X}\right)$, with $\phi$ representing the Galactocentric azimuth, measured positively in the direction of Galactic rotation.

The corresponding three-dimensional velocity of each star was derived following \cite{drimmel2023gaia}. First, in the heliocentric Cartesian coordinate system, the stellar velocity components ($u, v, w$) relative to the Sun were determined using the RGB stars' proper motions and line-of-sight velocities ($\mu_{\alpha^{*}}, \mu_{\delta}, v_{\rm los}$), where $\mu_{\alpha^{*}} = \mu_{\alpha} \cos\delta$. The velocity transformation is given by
\begin{equation}
V_{\text{rel}} = 
\begin{pmatrix}
u \\
v \\
w
\end{pmatrix}
= A_{G}' A
\begin{pmatrix}
4.74047\, \mu_{\alpha^{*}} d \\
4.74047\, \mu_{\delta} d \\
v_{\rm los}
\end{pmatrix}.
\end{equation}
Here, $ A_{G}' $ is a fixed orthogonal matrix that transforms equatorial coordinates to the Galactic frame, as defined in equation 4.62 of the \href{https://gea.esac.esa.int/archive/documentation/GEDR3/Data_processing/chap_cu3ast/sec_cu3ast_intro/ssec_cu3ast_intro_tansforms.html#SSS1}{$Gaia$ EDR3 online documentation}. The matrix $ A $ represents the local orthonormal basis at the star's position and is given by
\begin{equation}
A = 
\begin{pmatrix}
-\sin\alpha & -\sin\delta\cos\alpha & \cos\delta\cos\alpha \\
\cos\alpha & -\sin\delta\sin\alpha & \cos\delta\sin\alpha \\
0 & \cos\delta & \sin\delta
\end{pmatrix}.
\end{equation}

To express the stellar motion in a Galactocentric reference frame, we incorporated the solar motion vector and obtain $[u_*, v_*, w_*] = V_{\text{rel}} + [U_\odot, V_\odot, W_\odot]$, where we adopted the solar motion values $[U_\odot, V_\odot, W_\odot] = [11.1, 245.8, 7.8] \kms$ \citep{reid2004proper,reid2019_spiral}. Finally, we obtained the three-dimensional velocity components in Galactocentric cylindrical coordinates using the following transformations:
\begin{equation}
\begin{cases}
V_R = -u_*\cos\phi + v_*\sin\phi,\\
V_{\phi} = u_*\sin\phi + v_*\cos\phi,\\
V_Z = w_*.
\end{cases}
\end{equation}

To further refine the disk star sample, we implemented additional selection criteria, requiring that stars satisfy $ |V_Z| < 100 \kms$ and $ |Z| < 1 \kpc $. We also excluded the Galactic bar region by imposing an inner radial boundary of 
$ R > 5 ~\kpc$. These refined criteria yield a final sample of 46,330 RGB stars. 

We suppressed the observational noise and intrinsic statistical dispersion by applying a spatial smoothing to the radial velocity field in the X–Y plane. We first constructed a two-dimensional grid with an initial bin size of $dx=dy=0.25 \kpc$. Then we performed a smoothing operation for each bin individually. Specifically, for each bin center, we selected all stars within a square smoothing window of width 0.625 kpc (i.e., 2.5 times the initial bin size). Using the stars contained within each smoothing window, we applied the extreme deconvolution Gaussian mixture model (XDGMM) technique to fit a single-component Gaussian distribution, explicitly accounting for the individual measurement uncertainties of stellar radial velocities. This procedure yields one estimate of the mean radial velocity for the given window. To robustly quantify the statistical uncertainty, we generated 500 Bootstrap realizations of the stellar sample within each smoothing window. For each Bootstrap realization, the XDGMM fitting was performed once, resulting in a distribution of mean radial velocity estimates. The mean of this distribution was adopted as the radial velocity for this $0.25$-${\;\rm kpc}$ wide bin, while its dispersion was taken as the standard error of the mean (SEM). The resulting smoothed radial velocity map ($V_R$) and its corresponding SEM map ($\sigma_{V_R}$) are shown in Fig.~\ref{fig:xy_vr_sig}. In Panel (a), two red-blue transitions of the radial velocity $V_R$ are distinctly visible from small to large radii. The SEM $\sigma_{V_R}$ increases outward, which is from the combined effects of distance uncertainties and variations in the number of stars across different radial bins. The corresponding unsmoothed velocity maps are provided in the Appendix (Fig.~\ref{fig:ap_vrtz}) for reference.

In this work, we assume the coherent $V_R$ feature presented in Fig.~\ref{fig:xy_vr_sig} is caused by a steady-rotating spiral arm density wave in the Galactic disk. Therefore we can use it to constrain the properties of the underlying Galactic spiral potential, similar to the analysis in \citet{eilers2020strength}. We leave the possibility of a tidally induced spiral arm \citep[e.g.,][]{bernet_etal_25} to a future work. 

\section{Steady-state toy model}\label{sec:model}
\begin{table}
\centering
\caption{Parameters used in the perturbed spiral arm model.}
\begin{threeparttable}
\begin{tabular}{c c c} \hline
Symbol & Quantity & Value \\\hline
$\hr$ & scale length & $(1,50)\kpc$ \\
$\sigmamax$ & local peak surface density & $(1,30)\mpc$ \\
$p$ & pitch angle & $(5.7, 17.2)\deg$ \\
$t$ & time & $(6,9) \gyr$ \\
$m$  & wavenumber & 2 \\
$h_{z}$ & MW disk scale height & $1 \kpc$ \\
$V_{c}$ & circular velocity & $229 \kms$ \\
\hline
\multirow{2}{*}{$\op$\tnote{a}} & \multirow{2}{*}{pattern speed} & 2 and 12 $\kmskpc$ \\
 & & (1, 30) $\kmskpc$ \\ \hline 
\end{tabular}
\begin{tablenotes}
    \item[a] When the pattern speed is fixed, the model can avoid the Lindblad and corotation resonances. 
    Allowing $\Omega_p$ to vary within a broad prior introduces resonant effects.
\end{tablenotes}
\end{threeparttable}
\label{tab:model_fields}
\end{table}
We used epicyclic approximation to investigate the stellar velocity fields perturbed by a spiral potential \citep{binney2008galactic}. This is a classic approach that has been used in many studies \citep[e.g.,][]{schoenmakers1997measuring, siebert2012properties, eilers2020strength}. A detailed derivation is provided in Appendix \ref{sec:ap_mod}, while a brief outline is summarized here.

In this model, vertical velocity components are neglected, and the stellar motions are confined within the Galactic midplane. We adopted corotating polar coordinates $(R, \varphi)$, where $\varphi$ denotes the azimuth in the corotating frame, related to the inertial azimuth $\phi$ by $\varphi = \phi - \op t$. Here $\op$ represents the pattern speed of the spiral perturbation. The Galactic gravitational potential, $\Phi$, consists of an unperturbed axisymmetric component, $\Phi_0(R)$, and a non-axisymmetric perturbation, $\Phi_1(R, \varphi)$, expressed as  
\begin{align}
\Phi(R, \varphi) = \Phi_0(R) + \Phi_1(R, \varphi)
.\end{align}

We considered small deviations from a pure circular orbit at $ R=R_0 $, where the radial and azimuthal perturbations are denoted as $ R_1(t) $ and $ \varphi_1(t) $, respectively. The first-order terms of the equations of motion in the radial direction is
\begin{align}\label{eq:motion_R1}
\ddot{R}_1 + \left( \frac{\mathrm{d}^2\Phi_0}{\mathrm{d}R^2} - \Omega^2 \right)_{R_0} R_1 - 2 R_0 \Omega_0 \dot{\varphi}_1 
= -\left( \frac{\partial\Phi_1}{\partial R} \right)_{R_0},
\end{align}
while the equation of motion for the azimuthal direction is given by
\begin{align}\label{eq:motion_phi1}
\ddot{\varphi}_1 + 2 \Omega_0 \frac{\dot{R}_1}{R_0} = -\frac{1}{R_0^2} \left( \frac{\partial\Phi_1}{\partial\varphi} \right)_{R_0}.
\end{align}
Here, $\Omega_0$ represents the angular velocity of an unperturbed circular orbit. For the Milky Way, we assumed a flat circular rotation curve with a constant circular velocity of $ V_c(R) = 229 \, \text{km s}^{-1} $, which leads to an angular velocity of $ \Omega_0 = \frac{V_c}{R_0} $. If the  perturbation in the azimuthal direction is small, such that $\varphi_1 \ll 1$, the azimuthal coordinate of the orbit can be approximated as $\varphi \approx \varphi_0(t) = \phi_0 - \Omega_p t$, simplifying the analysis. 

We adopted a logarithmic form for the perturbing spiral potential given by  
\begin{align}\label{eq:spiralpot}
\Phi_1(R, \varphi) = \mathcal{A}(R) \exp \left(i \chi \right),
\end{align}
where \( \chi = m(\phi_{0}-\Omega_{p}t) + m\frac{\ln(R_{0}/h_{R,1})}{\tan p} \), $ m $ represents the azimuthal wavenumber (e.g., $ m = 2 $ for a two-armed spiral), $\hr$ is the radial scale length of the perturbation, and $ \mathcal{A}(R) $ denotes its amplitude. The parameter $ p $ characterizes the pitch angle of the spiral arms. Observational constraints suggest that $ p $ typically falls within the range of $ 10^\circ$–$20^\circ $ \citep{bobylev2014milky, vallee2017_p}, where the Wentzel-Kramers-Brillouin (WKB) approximation can be applied.

We then solved Equation \eqref{eq:motion_R1} and further derived the radial velocity, \( V_R = \dot{R}_1(t) \). We refer the reader to Appendix \ref{subsec:ap_vr_mod} for a detailed derivation. The final result is expressed as follows:
\begin{align}\label{eq:vr_full}
V_R &= \vsin + \vcos,
\end{align}
where
\begin{equation}
\label{eq:vr_full_sincos}
\begin{cases}
\vsin &= \left[\frac{2\Omega_{0}\mathcal{A}}{R_{0}(\Omega_{0}-\Omega_{p})}+\left(\frac{d\mathcal{A}}{dR}\right)_{R_{0}}\right]\frac{m(\Omega_{0}-\Omega_{p})}{\Delta}\sin \chi \\
\vcos &= \frac{m\mathcal{A}}{R_{0}\tan p}\frac{m(\Omega_{0}-\Omega_{p})}{\Delta}\cos\chi
\end{cases}
.\end{equation}
Here, the epicycle frequency is defined as \( \kappa_0^2 = \left(R \frac{\mathrm{d}\Omega^2}{\mathrm{d}R} + 4\Omega^2\right)_{R_0} \), leading to 
$\Delta = \kappa_0^2 - m^2(\Omega_0 - \Omega_p)^2$. The $\vsin$ model is obtained by considering only the real part of the perturbed potential (i.e., \citealt{eilers2020strength}), while the $\vcos$ term is an additional term derived from the imaginary part of the perturbed potential (see also \citealt{siebert2012properties}); we aim to demonstrate that both terms constitute important contributions to the velocity response of the perturbed spiral potential (see Figure~\ref{fig:ag_eq}). The perturbation amplitude, $\mathcal{A}(R)$, adopted in the above derivation follows the form proposed by \citealt{eilers2020strength}: 
\begin{align}
    \mathcal{A}(R) &= -\frac{2\pi G R^2 \tan^2 p}{h_z m^2}\Sigma_{\max}(R_{\odot})\exp\left(-\frac{R - R_{\odot}}{h_{R,1}}\right).
\end{align}
 Here, $\sigmamax$ represents the local maximum surface density of the spiral perturbation at solar radius, and its relation to the surface density profile is 
\begin{equation}\label{eq:sfd}
    \Sigma_1(R,\phi) = \Sigma_{\max}(R_{\odot})\exp\left(-\frac{R - R_{\odot}}{h_{R,1}}\right)\cos\chi.
\end{equation}

\section{Mock data}\label{sec:mock}

In this section, we present a comparison between our $\vr$ model and the orbit-integration results, and assess the viability of the Bayesian fitting approach. We subsequently discuss the physical interpretations of the individual fit parameters.

\subsection{Orbit integration}\label{subsec:agama}

\begin{figure*}
    \centering
    \includegraphics[width=\textwidth]{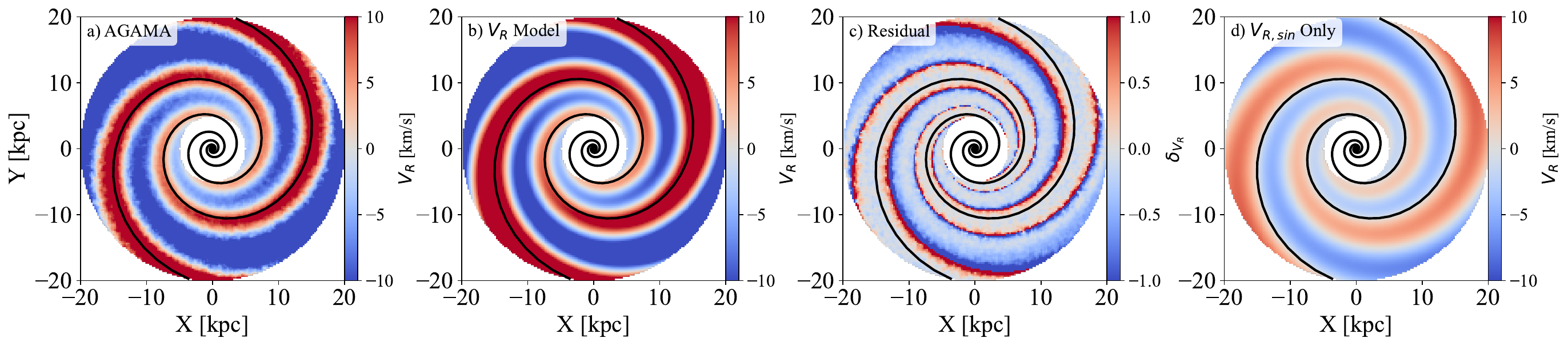}
    \caption{Comparison between the \texttt{AGAMA} orbit-integration results and our $\vr$ models. Panels (a-d) show, respectively, the mock $\vr$ field from \texttt{AGAMA}, the corrected $\vr$ model, their residuals ($\delta_{V_R}=\frac{V_{R,AG}-V_{R}}{V_{R}}$), and the $\vsin$ only model. Black spirals mark the loci of minimum gravitational potential.}
\label{fig:ag_eq}
\end{figure*}

\begin{table*}
\centering
\begin{threeparttable}
\renewcommand{\arraystretch}{1.5}
\caption{Parameters used in \texttt{AGAMA} and the corresponding fit parameters using the $\vr$ model.}
\label{tab:mock_result} 
\begin{tabular}{l c c c c c} 
\hline
Data Type & $\hr$ & $\sigmamax$ & $t$ & $p$ & $\chired$ \tnote{a}\\
\hline
Input Parameters & $11.36$ & $6$ & $7.47$ & $12.66$ & \\
Fitting result & $8.84^{+0.02}_{-0.02}$ & $5.83^{+0.01}_{-0.01}$ & $8.01^{+0.01}_{-0.01}$ & $12.60^{+0.01}_{-0.01}$ & $10.62$ \\
\hline
\end{tabular}
\begin{tablenotes}
     \item[a] $\chired = \frac{1}{n - k} \sum_{i=1}^{n} \frac{(V_{R,i} - V_{R_{\rm mod},i})^2}{|V_{R_{\rm mod},i}|}$ represents a modified reduced chi-square metric, where $k$ is the number of free parameters.
\end{tablenotes}
\end{threeparttable}
\end{table*}

We constructed the radial velocity map by directly performing orbit integrations in a composite gravitational potential. All orbit integrations were carried out using the \texttt{AGAMA} library \citep{agama2019}.

The background gravitational potential was modeled as a cored singular isothermal sphere evaluated in the Galactic midplane, and is given by
\begin{align}
\Phi_0 = -\frac{1}{2}V_{c}^2~\mathrm{log}(R^2+R_{c}^2)
.\end{align}
\citep{binney1981_logpot, evans1993_logpot}.
Here, $R_c=0.01\kpc$ is a small core radius introduced to regularize the potential near the Galactic center. This potential yields an approximately flat rotation curve with an asymptotic circular velocity of $V_c=229\kms$, and thus provides a simple and well-controlled background for the orbit integrations.

We additionally included an $m=2$ spiral-arm perturbation described by Eq.~\eqref{eq:spiralpot}. For the \texttt{AGAMA} implementation, the input parameter set was $\fullparam=$ $(11.36, 6, 7.47, 12.66)$ $\fullunit$ \citep{eilers2020strength}, and the orbits were integrated over a sufficient time ($\sim40\gyr$) to achieve a quasi-steady state. Figure~\ref{fig:ag_eq} presents an illustrative comparison of the orbit-integration results for $\op=2\kmskpc$ and the $\vr$ model. 
Panel~(a) displays the $\vr$ field obtained from the \texttt{AGAMA} orbit integration, and panel~(b) shows our $\vr$ model (Eq.~\ref{eq:vr_full}) map under the same parameter configuration. 
Panel~(c) presents the residuals between the two, showing deviations within approximately $\pm14\%$. The largest discrepancies occur primarily in regions where $\vr \approx 0\kms$, particularly in the outer disk, where the radial velocity field undergoes a blue-red transition and is therefore highly sensitive to small phase offsets between the linear model and the orbit integration. When these regions are excluded, the residuals decrease to about $2\%$. 
Panel~(d) illustrates only the $\vsin$ term of the model (Eq.~\ref{eq:vr_full_sincos}) for comparison. The dark spiral lines mark the loci of the local gravitational potential minima. In both the orbit integration and the $\vr$ model, these potential minima are generally located in regions of outward radial motion ($\vr > 0\kms$), whereas in the $\vsin$ model they coincide roughly with the blue-red transition zones where $\vr \approx 0\kms$. This comparison highlights the importance of including the $\vcos$ term to the radial-velocity response model.
 
Subsequently, we employed the Dynamic nested sampling (\href{https://dynesty.readthedocs.io/en/v3.0.0/}{\texttt{dynesty}\footnote{https://dynesty.readthedocs.io/en/v3.0.0}}) package \citep{dynesty, DYNESTY_Speagle, dynesty_sergey} to recover the input parameters of the spiral perturbation with a pattern speed of $\op = 2\kmskpc$. The likelihood function adopted in the fitting procedure is given by
\begin{align}
\mathcal{L} = \prod_{i=1}^{n} \frac{1}{\sqrt{2\pi\sigma_{V_R,i}^2}} 
\exp\left[-\frac{(V_{R,i} - V_{R_{\mathrm{mod}},i})^2}{2\sigma_{V_R,i}^2}\right],
\end{align}
where $V_{R,i}$ and $V_{R_{\mathrm{mod}},i}$ denote the mock and model-predicted radial velocities for each bin, and $\sigma_{V_R,i}$ represents represents the corresponding SEM estimated from the Bootstrap realizations\footnote{For each star particle, we assume a small $V_R$ uncertainty of $1.0\times10^{-6}\kms$ during the SEM calculation.}. The maximum-likelihood estimates are summarized in Table~\ref{tab:mock_result}. The recovered values of $\sigmamax$ and $p$ are nearly identical to the input parameters, whereas noticeable discrepancies are found in $\hr$ and $t$. These deviations primarily arise from systematic differences between the radial velocity field obtained from the \texttt{AGAMA} orbit integrations and that predicted by the analytic $\vr$ model, as illustrated in Fig.~\ref{fig:ag_eq}. In addition, both $\hr$ and $t$ are periodic and are strongly covariant in our model (see Eq.~\ref{eq:spiralpot} and Fig.~\ref{fig:ap_blindsearch}). While $t$ basically reflects the rotation angle of the spiral, $\hr$ additionally governs the amplitude variation (decay or growth) of $\vr$ along the radius. We find that $\hr$ recovered from the \texttt{dynesty} fit can be $\sim25\%$ smaller than its input value, possibly due to the fact that linear approximation described in Section~\ref{sec:model} is not accurate enough. The small fitting error represents the good convergence of the {\tt dynesty} procedure.

\subsection{Effects of pattern speed and resonances}
\label{subsec:res}

\begin{figure*}
    \centering
    \includegraphics[width=\textwidth]{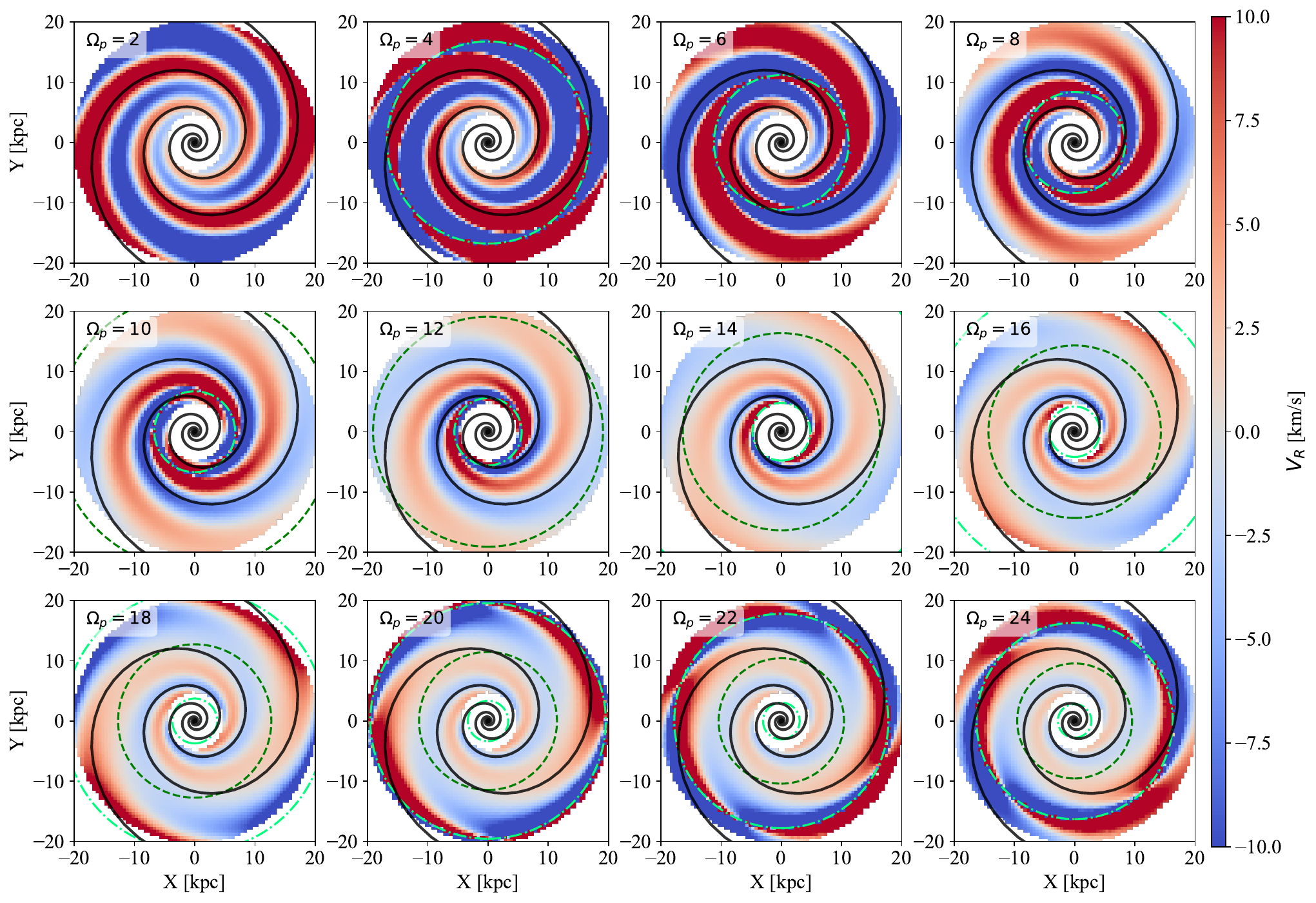}
    \caption{
    Results from the $\vr$ model with identical parameters but varying pattern speeds from $\op = 2$ to $24\kmskpc$ in steps of $2\kmskpc$. Dashed green circles mark the theoretical CR positions ($R_{\rm CR}$) for a flat rotation curve, while dash-dotted light green circles indicate the LR positions ($R_{\rm ILR}$ and $R_{\rm OLR}$). Black spirals denote the loci of the minimum gravitational potential.}
    \label{fig:vr_eq_res} 
\end{figure*}

\begin{figure}
    \centering
    \includegraphics[width=0.45\textwidth]{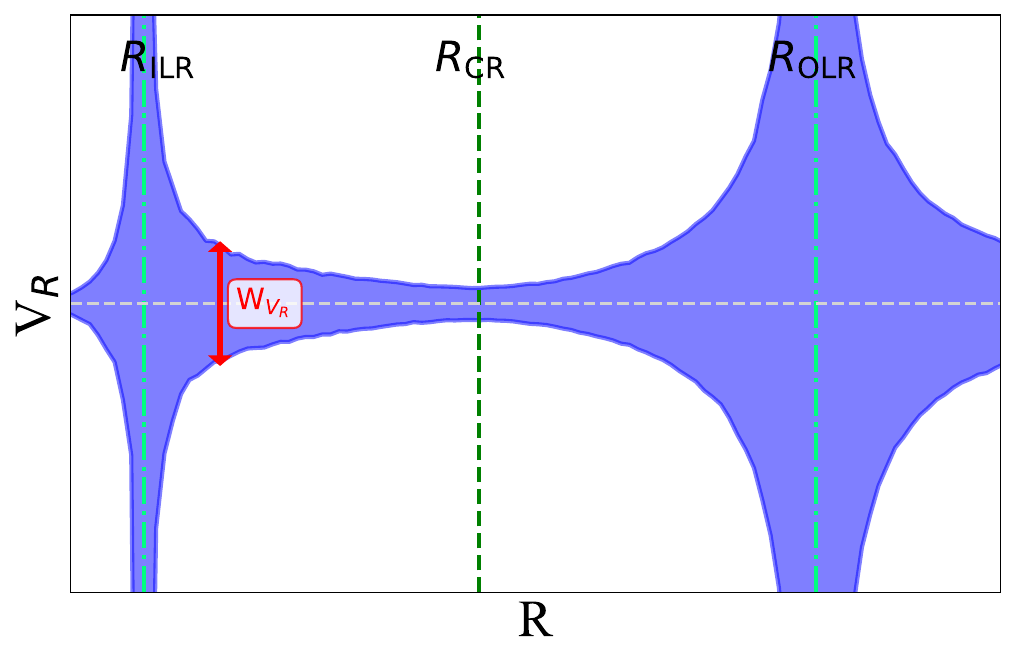}
    \caption{Schematic illustration of the relationship between the radial velocity field and resonance locations. The shaded blue region represents the distribution of $V_R$, vertical dashed lines indicate the resonance radii, and the horizontal dashed line marks $V_R = 0\kms$.}
    \label{fig:r_vr_res}
\end{figure}

\begin{figure*}
    \centering
    \includegraphics[width=\textwidth]{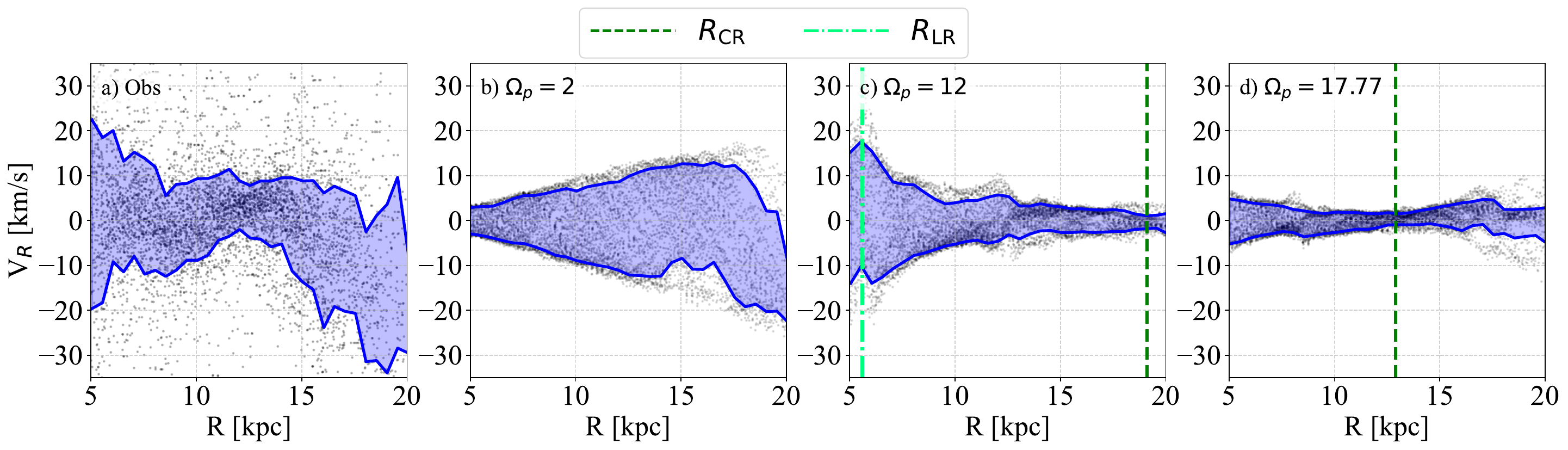}
    \caption{Radial velocity distribution as a function of Galactocentric radius after spatial binning. Panel (a) displays the observational distribution, while Panels (b-d) present the spatially resampled results derived from orbit integrations with $\op = 2, 12$, and $17.77\kmskpc$, respectively. The gray points represent the binned data points, the blue shaded regions denote the 16th and 84th percentile intervals.}
    \label{fig:r_vr_obs_ag}
\end{figure*}

In this section, we investigate the impact of the spiral pattern speed, $\op$, on the resulting $V_R$ maps using Eq.~\ref{eq:vr_full}. Since the parameters $\hr$ and $\sigmamax$ jointly regulate the overall amplitude of the radial velocity field, their effects are analyzed and presented separately in Appendix \ref{sec:ap_hr_sfd}. The parameter set adopted here remains $\fullparam=$ $(11.36, 6, 7.47, 12.66)$$\fullunit$, while $\op$ varies from 2 to $24\kmskpc$.

The results for different values of $\op$ are shown in Fig.~\ref{fig:vr_eq_res}. The solid black curves indicate the locations of the spiral arm potential minima, as in Fig.~\ref{fig:ag_eq}. For comparison purposes, we rotated the $\vr$ map in each panel such that all solid black curves are identical. The circles mark the theoretical resonance radii for a flat Galactic rotation curve with $V_c = 229\kms$. The dashed green lines denote the corotation resonance ($R_{\rm CR}$), while the dotted spring-green lines indicate the Lindblad resonances (LRs; $R_{\rm ILR}$ and $R_{\rm OLR}$). For $\op = 2\kmskpc$, $R_{\rm ILR} \approx 34\kpc$ such that our RGB sample ($5 < R < 20\kpc$) avoids all resonance regions. As $\op$ increases, the resonances progressively enter the observed radial range. For $\op = 4$--$10\kmskpc$, our sample region includes the inner Lindblad resonance (ILR). For $\op = 12$ and $14\kmskpc$, both the ILR and the corotation resonance (CR) are presented. For higher pattern speeds ($\op \gtrsim 20\kmskpc$), the CR and the outer Lindblad resonance (OLR) are encountered, while the ILR shifts inward to $R_{\rm ILR} < 5\kpc$, outside the radial coverage of the sample.

The resonances have large impacts on the radial velocity field. The LRs split the continuous logarithmic spiral pattern of $V_R$ into two separated parts with an abrupt phase transition (e.g., at $\op = 8$ and $24\kmskpc$ in Figure~\ref{fig:vr_eq_res}). As a consequence, the relative location of the spiral arms potential minima (or peak density) with respect to the radial velocity field changes systematically with $\op$. For $\op = 2\kmskpc$, the spiral arms are predominantly located in regions with $\vr > 0$ as they are inside $R_{\rm ILR}$. For $\op = 12\kmskpc$, they are instead associated with regions of $\vr < 0$ as they located between $R_{\rm ILR}$ and $R_{\rm CR}$. This would affect the predicted spiral arm density location when fitting to the real data (see Section~\ref{subsuc:previous}).

The amplitude of $V_R$ is also modulated with the resonances, as is illustrated schematically in Fig.~\ref{fig:r_vr_res}. The linear approximation fails at both the ILR and OLR, leading to sharp enhancements in $V_R$ (e.g., at $\op = 8$ and $24\kmskpc$). In contrast, the CR ($\op=\Omega_0$) is characterized by a significantly reduced radial velocity magnitude (e.g., at $\op = 18\kmskpc$). According to Eq.~(\ref{eq:vr_full_sincos}) the $\vsin$ component retains only its leading term (the divergent term in the linear approximation converges to a finite but small value), while the $\vcos$ component vanishes entirely ($\vcos=0\kms$) at $R=R_{\rm CR}$. To quantify the spread of $V_R$ as a function of radius, we defined the width of the $V_R$ distribution as ${\rm W}_{V_R}$, which is
\begin{align}
    {\rm W}_{V_R} = P_{84}(V_R|R) - P_{16}(V_R|R),
\end{align}
where $P_{84}(V_R|R)$ and $P_{16}(V_R|R)$ are the velocities corresponding to the 84th and 16th quartiles of the $V_R$ distribution at a given radius bin, $R$ (see Figure~\ref{fig:r_vr_res}). The radial gradient of ${\rm W}_{V_R}$ is written as $\nabla_R {\rm W}_{V_R}$. We observe that resonances induce sign reversals in $\nabla_R \rm{W}_{V_R}$: the gradient is positive for $R < R_{\rm ILR}$, negative for $R_{\rm ILR} < R < R_{\rm CR}$, positive for $R_{\rm CR} < R < R_{\rm OLR}$, and negative beyond $R_{\rm OLR}$.

The above analysis has shown the complexity when the fitting region contains resonances. We therefore first focus on two representative pattern speeds where the resonances are avoided, i.e., $\op = 2\kmskpc$ (inside $R_{\rm ILR}$) and $12\kmskpc$ (between $R_{\rm ILR}$ and $R_{\rm CR}$). As is detailed in Section~\ref{sec:result}, we also explore a model in which $\op$ is treated as a free parameter, for which the best-fitting value we obtained is $\op = 17.77\kmskpc$. We performed orbit-integration simulations using {\tt AGAMA} to compare these three pattern speeds directly with the observed radial distribution of $\vr$. The input parameters for the potential are listed in Table~\ref{tab:result_apo}.

The results are shown in Fig.~\ref{fig:r_vr_obs_ag}. To make a fair comparison, the simulated star particles were sampled to match the spatial distribution of the observational data, and the same spatial smoothing procedure was applied. Consistent with the predictions of our $V_R$ model, the simulations show that the particles located near the LRs exhibit large $\vr$ amplitudes, whereas particles near the CR are characterized by significantly smaller radial velocities. In the observational data, ${\rm W}_{V_R}$ decreases with increasing radius for $R \lesssim 13\kpc$, suggesting that spiral-driven radial streaming motion weakens from the inner to the outer disk along the spiral arms. This behavior is consistent with the orbit-integration results for $\op = 12\kmskpc$ (Panel~c). In contrast, for $R \gtrsim13\kpc$, the observed ${\rm W}_{V_R}$ expands with increasing radius, in agreement with the simulation adopting $\Omega_p = 2\kmskpc$ (Panel~b). The decrease-increase trend of ${\rm W}_{V_R}$ in the observation is better reproduced by the model with $\Omega_p = 17.77\kmskpc$ (Panel~d), where $R_{\rm CR}$ roughly lies at $13\kpc$ (i.e., the turning point of the ${\rm W}_{V_R}$ profile). However, we note in this case the typical spiral-driven radial velocity remains relatively modest, with $|V_R| \lesssim 10\kms$.

\section{Kinematic signature of spiral arms}\label{sec:result}
\begin{table*}
\centering
\renewcommand{\arraystretch}{1.5}
\begin{threeparttable}
\caption{Best-fit parameters for non-axisymmetric spiral arm models.}
\label{tab:result_apo}
\begin{tabular}{c c c c c c c c c}
\hline
Model & Fitting region & $\op$  & $\hr$ & $\sigmamax$ & $t$ & $p$ & $\chired$ & $\xi$ \tnote{a}\\
\hline 
\multirow{2}{*}{Fixed $\op$} 
& 5-20 & 2  & $\gtrsim 50$ & $2.42^{+0.05}_{-0.05}$ & $7.20^{+0.05}_{-0.03}$ & $10.89^{+0.01}_{-0.01}$ & $149.08$ & $3.6\%$\\ \cline{2 - 9}
&  7-18  & 12 & $\gtrsim 50$ & $4.43^{+0.08}_{-0.08}$ & $7.29^{+0.04}_{-0.01}$ & $12.60^{+0.01}_{-0.01}$ & $173.35$ & $6.5\%$\\ 
\hline
Free $\op$ & 5-20 & $17.77^{+0.05}_{-0.02}$ & $39.26^{+10.74}_{-7.43}$ & $12.38^{+0.51}_{-0.43}$ & $7.56^{+0.61}_{-1.48}$ & $8.02^{+0.01}_{-0.01}$  & $138.81$ & $17.8\%$ \\
\hline
\end{tabular}
\begin{tablenotes} 
    \item[a] $\xi=\frac{\Sigma_1}{\Sigma_0}|_{R_{\odot}}$ represents the local maximum surface density contrast at solar radius, where $\Sigma_0 =68 \mpc$ is the stellar disk surface density at the solar radius \citep{bovy2013_background}.
\end{tablenotes}
\end{threeparttable}
\end{table*}

\begin{figure*}
    \centering
    \includegraphics[width=\textwidth]{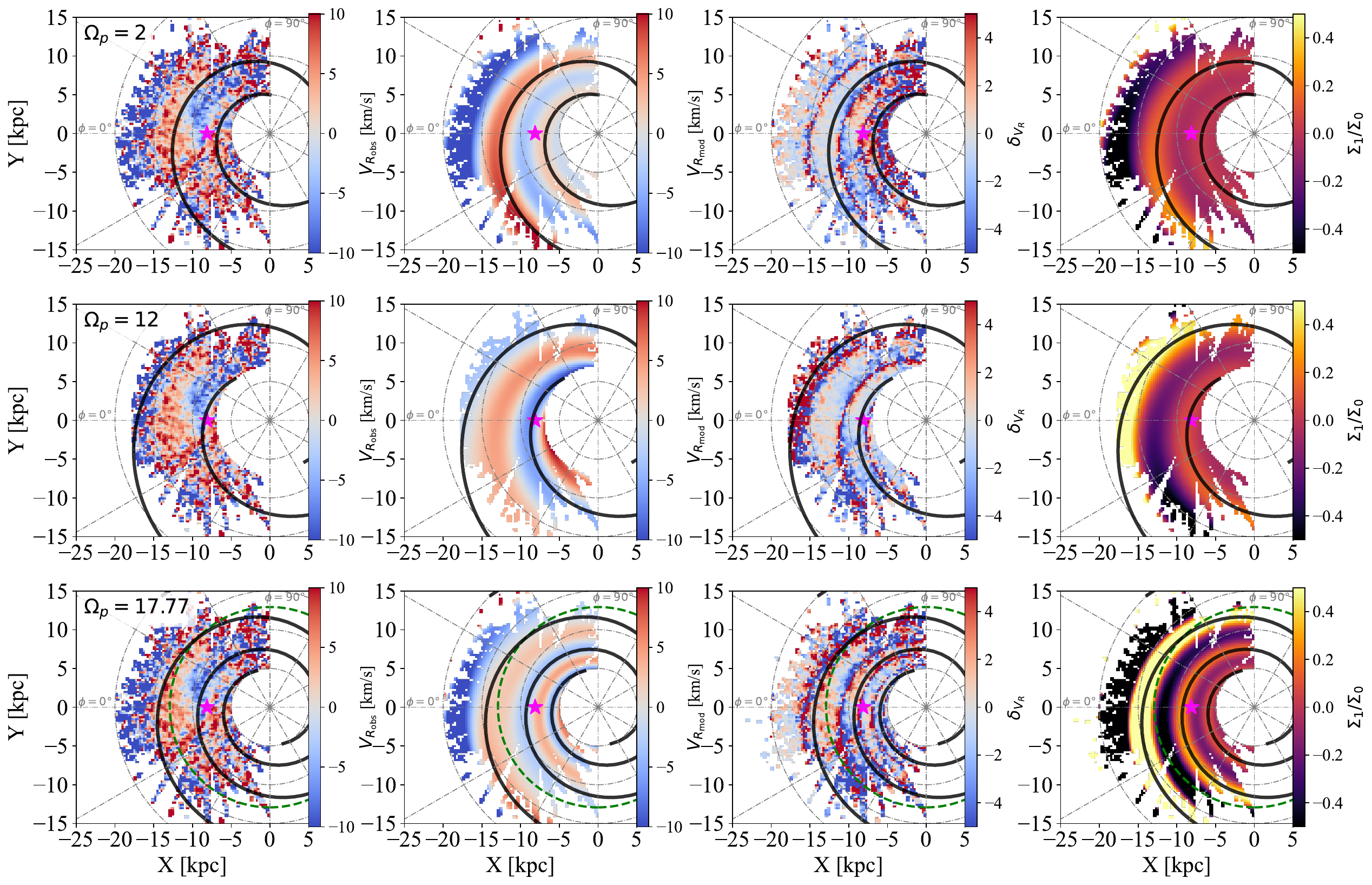}
    \caption{Spiral-arm loci traced by the RGB sample kinematics. The first and second rows present the results for fixed pattern speeds of $\op = 2$ and $12\kmskpc$, respectively, while the third row shows the result for a free pattern speed with $\op = 17.77\kmskpc$. From left to right, the columns display the observational velocity maps, the best-fit model maps, the residual maps, and the stellar surface density contrast maps. Solid black curves denote the fit spiral-arm loci, and the green line marks the CR.}
\label{fig:result_apo}
\end{figure*}

We proceeded to fit the $V_R$ model to the APOGEE DR17 and $\gaia$ DR3 RGB sample, excluding the highly noisy region with $X > 0\kpc$. We performed three tests with different spiral pattern speeds, $\op$: two with fixed values, $\Omega_p = 2$ and $12\kmskpc$, following \citet{eilers2020strength}, and one in which $\Omega_p$ was treated as a free parameter. The results are summarized in Table~\ref{tab:result_apo} and Fig.~\ref{fig:result_apo}.

To avoid the influence of resonances, for the fixed representative values $\op = 2$ and $12\kmskpc$, the fitting regions were restricted to $R = 5$--$20~\kpc$ (entirely within the ILR) and $R = 7$--$18\kpc$ (between the ILR and the CR), respectively. Figure~\ref{fig:result_apo} illustrates a comparison between the dataset and the best-fit models, where the fit spiral-arms loci are marked with black lines. As was discussed in the previous section, for $\op = 2\kmskpc$ (top row) the model predicts a positive $\nabla_R {\rm W}_{V_R}$ (second column), which leads to larger residuals in the inner disk (third column). In contrast, for $\op = 12\kmskpc$ (second row) the fit region already crosses the ILR, where the model exhibits a decreasing radial velocity gradient and larger residuals emerge in the outer disk. In addition, in the observational data the inner-disk signal appears more irregular and tightly wound, which further complicates the fitting for the $\op = 12\kmskpc$ model. As a result, its reduced chi-square value, $\chired$, is slightly larger than that obtained for $\op = 2\kmskpc$ (see Table ~\ref{tab:result_apo}). The fit spiral-arm loci coincide with the maxima of the stellar surface density contrast (see forth column); however, the arms traced by the $\op = 2\kmskpc$ model generally show a weaker contrast and are located closer to the Galactic center.

For the case in which $\op$ is treated as a free parameter (with a uniform prior of $\op = 1$--$30\kmskpc$), no additional restriction is imposed on the fitting region ($R = 5$--$20\kpc$). As is shown in the third row of Fig.~\ref{fig:result_apo}, this model inevitably encounters the CR at $R \approx 13\kpc$, where the predicted $|V_R|$ becomes very small, slightly underestimating the observed values. On either side of $R_{\rm CR}$, however, the model is able to simultaneously reproduce the decreasing trend of the maximum $V_R$ amplitude in the inner disk and the increasing trend in the outer disk. As a result, this model yields the smallest $\chi^2_{\rm red}$. In addition, the outermost spiral arm in this model appears to be a continuation of the inner spiral arm through azimuthal winding, and the corresponding stellar surface density contrast map exhibits the most prominent spiral-arm features.

\section{Discussion}\label{sec:dis}
\subsection{Comparison with previous studies}\label{subsuc:previous}
\begin{figure*}
    \centering
    \includegraphics[width=0.8\textwidth]{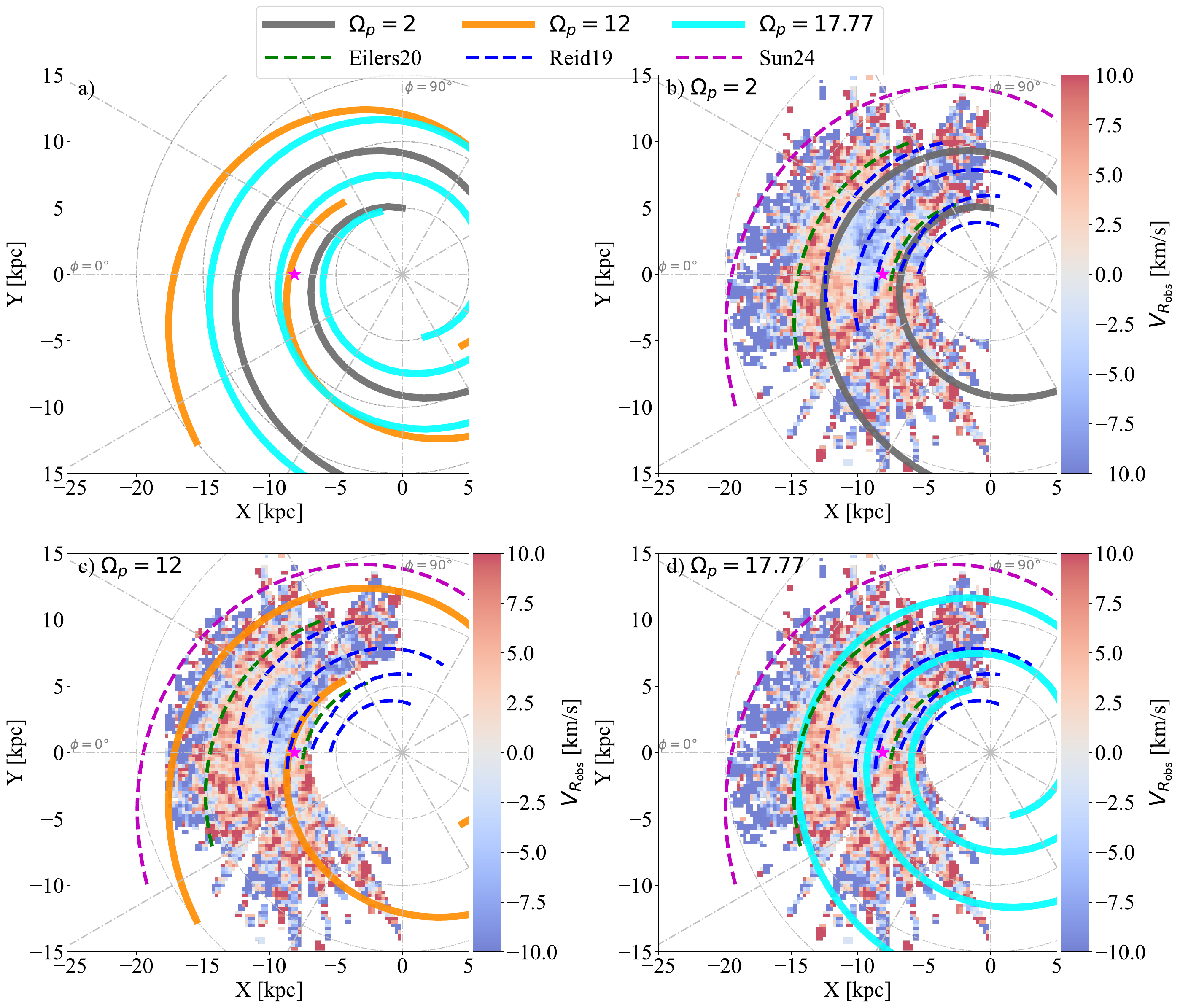}
    \caption{Comparison between the kinematically predicted spiral-arm features and previous studies. Panel (a) shows the three spiral-arm patterns derived in this work. Panels (b-d) compare our results with the spiral-arm loci from \citet{eilers2020strength} (thin dashed green line), \citet{reid2019_spiral} (thin dashed blue line), and \citet{sun2024_osc} (thin dashed magenta line), overlaid on the background map of mean radial velocity. From outer to inner, the thin dashed magenta and blue lines correspond to the Outer Scutum–Centaurus (OSC) arm, Outer Norma arm, Perseus arm, Local arm, Sagittarius–Carina arm, and Scutum–Centaurus arm.}
\label{fig:sa_all}
\end{figure*}

First, we discuss the four free parameters describing the perturbed potential. The derived radial scale lengths, $h_r$, range from $40$ to $50~\kpc$, consistent with the values reported by \citet{eilers2020strength} ($11$--$50~\kpc$). In our tests with a fixed spiral pattern speed, the fits tend to favor larger values of $h_r$ (e.g., $\gtrsim 50~\kpc$). This preference arises because the observed ${\rm W}_{V_R}$ exhibits a non-monotonic behavior, decreasing at smaller radii and increasing again in the outer disk (see Fig.~\ref{fig:r_vr_obs_ag}). When adopting a fixed $\op = 2$ or $12\kmskpc$, the model enforces a purely increasing or decreasing gradient, which can only be accommodated by increasing $h_r$ such that the resulting gradient becomes nearly flat. 
For the fixed-pattern-speed models, the fit local surface density contrast is relatively weak ($\xi \sim 5\%$), slightly below the $\sim 10\%$ reported by \citet{eilers2020strength} for a $\vsin$-only model. Because our model retains both the $\vsin$ and $\vcos$ terms, which represent important components of the perturbed potential, it requires a smaller perturbation amplitude $\sigmamax$, implying a lower overall surface density contrast in the Galactic stellar disk. When the pattern speed is treated as a free parameter ($\op = 17.77\kmskpc$), the fit yields a more moderate radial scale length, $\hr \approx 40~\kpc$, and a surface density contrast amplitude $\xi \simeq 17.8\%$, in closer agreement with previous studies \citep{khalil2024_vel, siebert2012properties}. However, in this case the fit inevitably encounters the CR, which tends to enhance the inferred $\sigmamax$.
In addition, the fit pitch angles cluster around $p \approx 10^\circ$, consistent with earlier measurements \citep{bobylev2014milky, vallee2017_p}. The spiral-arm rotation time, $t$, also cannot be reliably constrained within our framework, as the model assumes a steady-state spiral structure and is therefore insensitive to the absolute evolutionary phase.

In Figure~\ref{fig:sa_all}, we compare the spiral-arm signatures predicted by our radial velocity models with those reported in the literature. 
Panel~(a) shows the locations of the main spiral arms predicted by our three models. Within the observational coverage of our data, only two spiral arms are visible for $\op = 2$ and $12\kmskpc$, whereas for $\op = 17.77\kmskpc$ three arms are apparent, one of which originates from the winding of an inner arm into the outer disk. 
Panels~(b--d) further compare our predictions with spiral-arm loci derived from previous studies. \citet{reid2019_spiral} and \citet{sun2024_osc} constrained spiral-arm positions using high-precision maser and carbon monoxide (CO) observations, respectively. \citet{eilers2020strength}, similar to our approach, inferred spiral-arm locations kinematically from the $V_R$ map but based solely on a $\vsin$-only model. 
Our results show that for $\op = 2\kmskpc$, the predicted main-arm locations are broadly consistent with those reported by \citet{eilers2020strength}. In this case, the primary arms are likely associated with the Outer--Norma arm and the Sagittarius--Carina arm. For $\op = 12\kmskpc$, our outer arm lies in the middle of the Outer Scutum--Centaurus (OSC) arm and the Outer--Norma arm, while the inner arm corresponds to the Local arm. For the free-pattern-speed fitting case, $\op = 17.77\kmskpc$, the two dominant spiral arms are located near the Perseus arm, and between the Outer--Norma and OSC arm.

\subsection{Possible range of spiral pattern speeds}\label{subsec:op}
\begin{figure}
    \centering
    \includegraphics[width=0.4\textwidth]{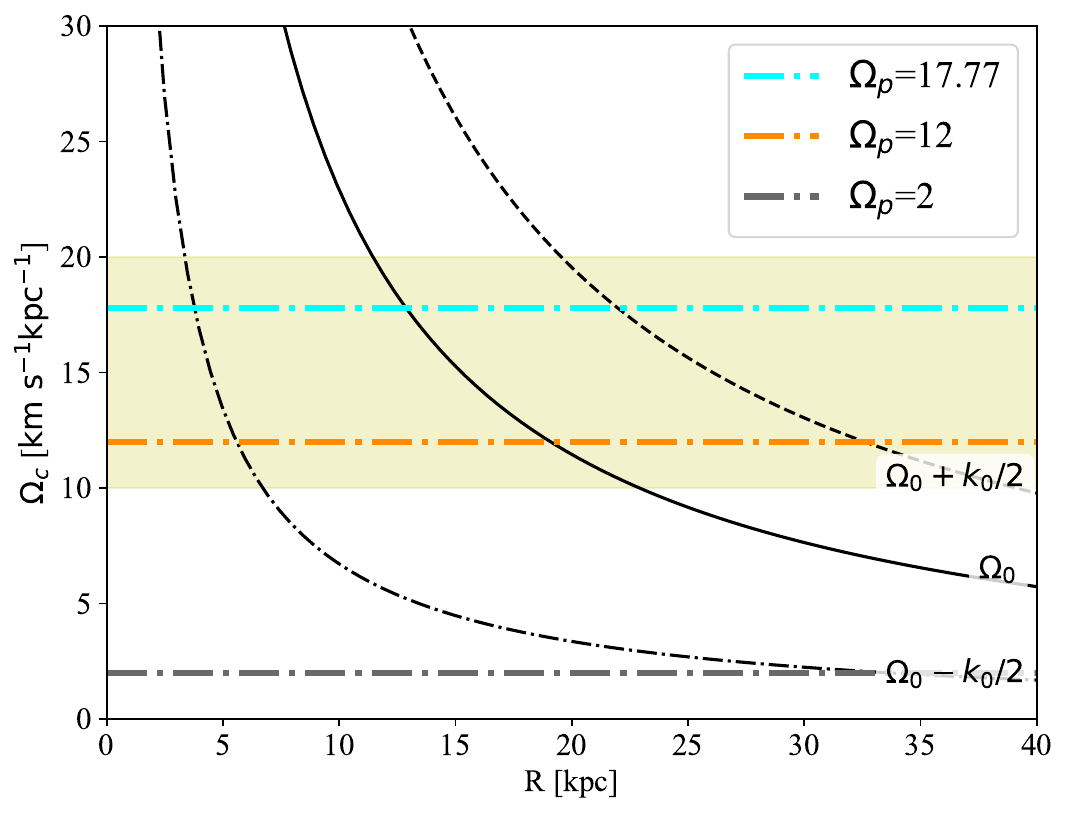}
    \caption{Corresponding frequency curves and resonances in our model. The shaded region indicates the range of spiral pattern speeds inferred from our $V_R$ model.}
\label{fig:frequency_curve}
\end{figure}

\begin{table} 
\centering
\begin{threeparttable}
\renewcommand{\arraystretch}{1.5}
\caption{Spiral pattern speeds allowed by the full observational coverage (including resonance effects).} \label{tab:result_op}
\begin{tabular}{ c c c c c c}
\hline
$\op$  & $\hr$ & $\sigmamax$ & $t$ & $p$ & $\chired$ \\
\hline 
5  & $\lesssim 1$ & $\lesssim 1$ & $7.30^{+0.07}_{-0.07}$ & $0.17$ & $1199329$ \\ 
10 & $\gtrsim 50$ & $\lesssim 1$ & $6.61^{+0.03}_{-0.01}$ & $\lesssim 0.1$ & $856$ \\ 
15 & $\gtrsim 50$ & $11.46^{+0.31}_{-0.30}$ & $7.5^{+0.03}_{-0.01}$ & $0.15$ & $202$ \\ 
20 & $2.64^{+0.08}_{-0.07}$ & $6.34^{+0.71}_{-0.67}$ & $7.47^{+0.01}_{-0.01}$ & $0.15$ & $425$ \\ 
25 & $2.11^{+0.38}_{-0.16}$ & $2.68^{+0.94}_{-0.91}$ & $7.23^{+1.61}_{-0.26}$ & $0.11$ & $2833$ \\ 
30 & $1.32^{+0.04}_{-0.07}$ & $2.74^{+0.58}_{-0.23}$ & $6.01^{+0.11}_{-0.01}$ & $0.13$ & $102241$ \\ 
\hline
\end{tabular}
\end{threeparttable}
\end{table}

We further explored the range of spiral pattern speeds that may be allowed within the current observational coverage when resonance effects are explicitly taken into account. To this end, we performed a simple exploratory test. Since an extremely slow pattern speed, such as $\op = 2\kmskpc$, is difficult to reconcile with most theoretical expectations, we restricted the tested range to $\op = 5$--$30\kmskpc$, sampled in steps of $5\kmskpc$. For each value of $\op$, we fit the $V_R$ model over the full observational domain. The fitting results are summarized in Table~\ref{tab:result_op}. By combining these results with the frequency curves of the background potential (Fig.~\ref{fig:frequency_curve}), we find that it is generally difficult for the model to avoid the nonlinear perturbations induced by the inner and outer LRs. For pattern speeds that place the LRs well within the observed radial range (e.g., $\op = 5$ or $30\kmskpc$), the fit $V_R$ maps exhibit very large amplitudes only within narrow regions near the LRs, while the radial velocity signal is strongly suppressed elsewhere. As a consequence, these cases yield very large values of $\chired$. If the presence of the CR is allowed, the overall fit quality improves significantly. This behavior suggests that the spiral pattern speed is most likely constrained to the range $\op \approx 10$--$20\kmskpc$.

\subsection{Limitations and future work}

In this work, we restrict ourselves to the classical quasi-steady density wave scenario, similarly to \citet{eilers2020strength}. We acknowledge several important limitations.
First, the model assumes that spiral arms can be approximated as long-lived, rigidly rotating global structures. However, both theoretical and numerical studies suggest that spiral arms can also be short-lived and recurrent \citep{Dobbs2014_transient_theory, Sellwood2022_transient_theory}, especially in the Milky Way\citep[][]{Hunt2025_tidal}. In this context, the global pattern speed derived in this work should be interpreted as an effective, time-averaged, or locally defined quantity, rather than evidence of a strictly steady global mode.

Second, the observed $V_R$ spiral features are not perfectly continuous and/or globally symmetric, which makes it difficult for the model to simultaneously reproduce all observed structures within a single, coherent solution. While the red–blue transition is prominent in the $Y > 0$ region, the inward-streaming signals (blue, $V_R < 0$) in the $Y < 0$ region do not smoothly connect with their counterparts at positive $Y$. Although this asymmetry may partly reflect limitations in signal-to-noise or spatial sampling, it may also indicate that a rigid, symmetric grand-design spiral model is an oversimplified representation of a more complex, possibly fragmented or transient structure.

Third, the present analysis focuses on two-dimensional in-plane kinematics ($V_R$ and $V_\phi$) and does not incorporate vertical ($Z$-direction) phase-space information. It is now well established that the Milky Way disk is in a state of disequilibrium \citep{Antoja2022_tidal, Hunt2025_tidal}, with external perturbations in the vertical direction. The coupling between in-plane spiral perturbations and vertical kinematic features is therefore not captured within our current 2D steady-state framework.

In future work, we plan to relax the quasi-steady assumption by incorporating time-dependent and transient gravitational potentials. Extending the analysis to include vertical kinematics will be essential for disentangling internal spiral dynamics from externally driven perturbations, ultimately enabling a more complete three-dimensional dynamical characterization of the Milky Way’s spiral structure.

\section{Summary} \label{sec:summary}
Based on the mean radial velocities of APOGEE DR17 and $\gaia$ DR3 RGB stars, we investigated the spiral signatures of the Milky Way using a revised $V_R$ model that incorporates both the $\vsin$ and $\vcos$ components. The model describes non-axisymmetric perturbations under the assumption of a two-armed spiral pattern.

To validate the revised model and the feasibility of the fitting procedure, we performed orbit integrations using \texttt{AGAMA} over a duration of $\sim40\,\gyr$. The revised model successfully reproduces the phase of the radial velocity field predicted by the orbit integrations. After excluding particles located near the transition regions between inward and outward radial motions, where the linear approximation is intrinsically unstable, the mean discrepancy in the $V_R$ amplitude is reduced to $\sim2\%$. This represents a substantial improvement over the original $\vsin$-only model.

We then used the \texttt{AGAMA}-generated mock dataset and applied Bayesian inference with \texttt{dynesty} to recover the input parameters. The results demonstrate that the local surface-density contrast $\sigmamax$ and the pitch angle, $p$, can be reconstructed. However, the parameters $\hr$ and $t$ exhibit strong covariance because both are periodic in our model. Consequently, $\hr$ is recovered with an uncertainty of $\sim 20\%$, which is effectively irreducible within the model. We also examined the influence of resonances and pattern speed on the radial velocity. The LRs cause strong deviations where linear theory breaks down, while the CR suppresses the radial-velocity signal. As the stellar location crosses each resonance, both the sign of $\nabla_R {\rm W}_{V_R}$ and the phase of the spiral-induced velocity pattern reverse. For example, for $\op = 2\kmskpc$, we obtain $\nabla_R {\rm W}_{V_R} > 0$, and the spiral-arm signal appears in regions with $V_R > 0$. In contrast, for $\op = 12\kmskpc$, we find $\nabla_R {\rm W}_{V_R} < 0$, and the spiral-arm signal is located in regions with $V_R < 0$. For $\op \simeq 18\kmskpc$, the presence of the CR at $R \approx 13\,\kpc$ leads to a mixed behavior: at $R < 13\,\kpc$ the radial velocity gradient and the spiral-arm phase resemble those of the $\op = 12\kmskpc$ case, while at $R > 13\,\kpc$ they are similar to the $\op = 2\kmskpc$ case, albeit with a smaller $V_R$ amplitude overall.

Using the same methodology, we fit the smoothed two-dimensional radial-velocity map from APOGEE and $\gaia$. For the fixed-pattern-speed cases ($\op = 2$ and $12\kmskpc$), we masked the bulge and resonance-dominated regions and adopted fitting ranges of $R = 5$–$20\kpc$ and $R = 7$–$18\kpc$, respectively. In contrast, for the free-pattern-speed model, no additional masking of resonance regions was applied.
First, the pitch angle remains close to $p \simeq 10^\circ$, consistent with previous studies \citep{bobylev2014milky}.
Second, the revised $\vr$ model reduces the implied density-perturbation amplitude by roughly a factor of two compared to earlier kinematic models for fixed pattern speeds, yielding a surface-density contrast of $\sim5\%$ at the solar radius. When the pattern speed is treated as a free parameter, the fit instead favors a higher contrast, $\xi \simeq 18\%$, consistent with previous studies.
Third, the inferred radial scale length of the perturbation depends on the assumed pattern speed: for the fixed pattern-speed cases, the fits favor large values of $\hr \gtrsim 50~\kpc$, whereas allowing $\op$ to vary freely yields a more moderate scale length, $\hr \approx 40~\kpc$. This behavior reflects parameter degeneracy driven by the non-monotonic radial profile of ${\rm W}_{V_R}$.

The inferred locations of the primary spiral arms depend on the assumed pattern speed. For $\Omega_p = 2\kmskpc$, the arms are consistent with the Sagittarius--Carina arm and the Outer--Norma arm, whereas for $\Omega_p = 12\kmskpc$, they shift slightly outward to align with the Local arm and an arm between the Outer--Norma and OSC arms. In the free-pattern-speed case ($\Omega_p = 17.77\kmskpc$), the dominant spiral arms are located further outward, corresponding to the Perseus arm and an arm between the Outer--Norma and the OSC arm.

Finally, by fitting the $V_R$ model over a range of pattern speeds while accounting for resonance effects, we find that pattern speeds placing the LRs within the observed radial range produce strongly localized and unrealistic velocity signatures, leading to poor fits. Allowing for the presence of corotation significantly improves the fit quality, indicating that the spiral pattern speed is most likely constrained to $\op \approx 10$--$20\kmskpc$. Future datasets with better spatial coverage and higher precision will allow for a more comprehensive three-dimensional characterization of the Milky Way’s dynamical spiral structure.

\begin{acknowledgements}
We are grateful to Juntai Shen, Sergey Koposov and Jing Li for helpful discussions and constructive comments that greatly benefited this study. This research is supported by the science research grants from the China Manned Space Project (No.CMS-CSST-2025-A11), the Innovation Program of Shanghai Municipal Education Commission (Grant No. 2025GDZKZD04), and the Program of China Scholarship Council (Grant No. 202608310096). I.T.S. acknowledges support from the Natural Science Foundation of China (Grant No. 12203035). H.T is supported by National Key R\&D Program of China No. 2024YFA1611902. Z.L. acknowledges the support of the National Natural Science Foundation of China (Grant No. 12573009) and the scientific research grants from the China Manned Space Project with Grant No. CMS-CSST-2025-A07 and CMS-CSST-2025-A05. Funding for the Sloan Digital Sky Survey IV has been provided by the Alfred P. Sloan Foundation, the U.S. Department of Energy Office of Science, and the Participating Institutions. This work has made use of data from the European Space Agency (ESA) mission {\it Gaia} (\url{https://www.cosmos.esa.int/gaia}), processed by the {\it Gaia} Data Processing and Analysis Consortium (DPAC, \url{https://www.cosmos.esa.int/web/gaia/dpac/consortium}). Funding for the DPAC has been provided by national institutions, in particular the institutions participating in the {\it Gaia} Multilateral Agreement.

\end{acknowledgements}

\bibliographystyle{aa}
\bibliography{sa_bib}

\begin{appendix}

\onecolumn
\section{Unsmoothed 3D velocity maps} \label{sec:vrtz}
\setcounter{figure}{0}
\renewcommand{\thefigure}{A\arabic{figure}}
\begin{figure*}[htp!]
    \centering
    \includegraphics[width=\textwidth]{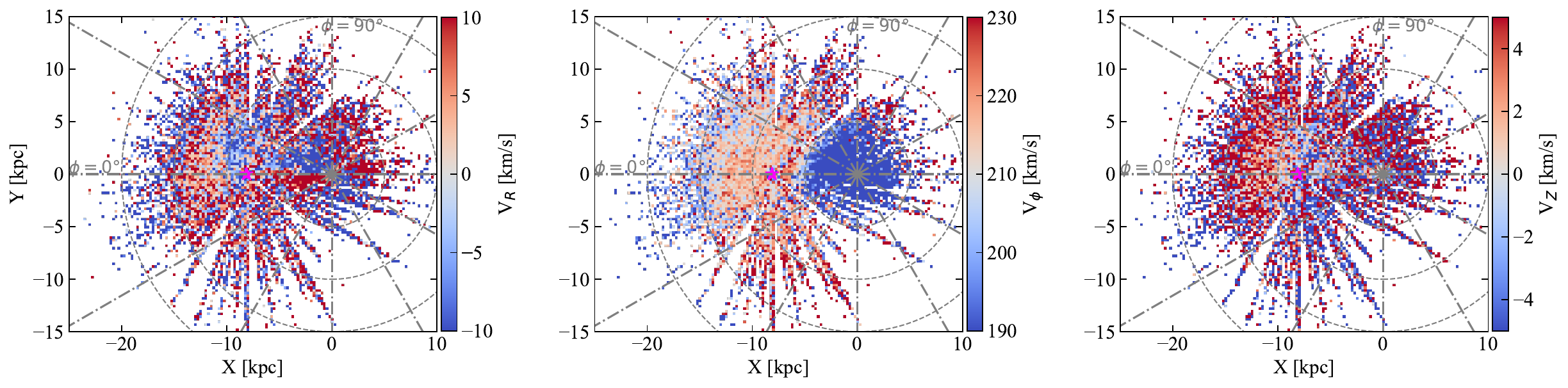}
    \caption{unsmoothed three-dimensional velocity maps of RGB stars. From left to right, the panels represent the radial, azimuthal, and vertical velocity components.}
\label{fig:ap_vrtz} 
\end{figure*}

Figure~\ref{fig:ap_vrtz} presents the unsmoothed three-dimensional velocity maps.
Compared with the smoothed maps used in the main analysis, the unsmoothed distributions exhibit stronger pixel–to–pixel noise and reduced spatial coherence, particularly in the outer disk where the number of stars per bin decreases. The applied smoothing procedure not only suppresses statistical noise but also enhances the visibility of large-scale velocity signals.

In principle, the azimuthal and vertical velocity components can also be used to trace non-axisymmetric perturbations in the Galactic disk. However, as shown in Figure~\ref{fig:ap_vrtz}, the spiral-related signatures in these components are significantly weaker than those in the radial velocity. Owing to distance limitations in our sample, only a single red–blue transition feature is visible in both the azimuthal and vertical velocity maps. Future datasets with improved spatial coverage and distance precision will be essential for a more detailed investigation of three-dimensional kinematic perturbations in the Galactic disk.

\section{Dynamical spiral perturbation toy model} \label{sec:ap_mod}
\subsection{Perturbed radial velocity model} \label{subsec:ap_vr_mod}
We model the orbits of stars in the presence of a weak perturbation potential, following the approach of \cite{binney2008galactic} (e.g., Section 3.3.3). The general loop orbits are described as a superposition of the guiding center motion and small oscillations around it, analogous to the epicycle theory for nearly circular orbits in an axisymmetric potential. Vertical velocity components are neglected, and the kinematics are considered within the Galactic midplane. 

Using polar coordinates $(R, \varphi)$ in the corotating frame of the spiral, the angular position for a retrograde orbit is defined as $\phi = \varphi + \Omega_p t$, where $\Omega_p$ is the pattern speed. The Lagrangian for the system is:
\begin{equation}
\mathcal{L}=\frac{1}{2}\dot{R}^2+\frac{1}{2}\left[R\left(\dot{\varphi}+\Omega_b\right)\right]^2-\Phi(R,\varphi),
\end{equation}

So the equations of motion in this frame are:
\begin{equation}\label{eq:a_R_phi}
\begin{cases}
    \ddot{R} = R\left(\dot{\varphi} + \Omega_p\right)^2 - \frac{\partial \Phi}{\partial R},\\ 
    \frac{\mathrm{d}}{\mathrm{d}t}\left[R^2\left(\dot{\varphi} + \Omega_p\right)\right] = -\frac{\partial \Phi}{\partial \varphi},
\end{cases}
\end{equation}

Consider an axisymmetric potential $\Phi_0(R)$. Assuming that it is perturbed by a weak spiral, the total potential can be expressed as:
$\Phi(R, \varphi) = \Phi_0(R) + \Phi_1(R, \varphi)$, where $\left|\frac{\Phi_1}{\Phi_0}\right| \ll 1$. Under this perturbation, the motion of a star can be described by decomposing $R$ and $\varphi$ into zero - order and first - order components: $R(t) = R_0 + R_1(t), \varphi(t) = \varphi_0(t) + \varphi_1(t)$. Here, $R_0$ is the radius of a circular orbit in the unperturbed axisymmetric potential, and $\varphi_0(t)$ is the angular position in the rotating frame. The angular velocity of the guiding center $(R_0, \varphi_0)$ is given by $\dot{\varphi}_0 = \Omega_0 - \Omega_p$, where $\Omega_0 \equiv \sqrt{\frac{1}{R_0} \frac{\mathrm{d} \Phi_0}{\mathrm{d} R}}$ is the angular velocity in the unperturbed potential. Thus, first-order terms in the equations of motion can be rewritten as:
\begin{subequations}\label{eq:apd_motion1}
\begin{empheq}[left={\empheqlbrace}]{align}
& \ddot{R}_1 + \left(\frac{\mathrm{d}^2\Phi_0}{\mathrm{d}R^2} - \Omega^2\right)_{R_0}R_1 - 2R_0\Omega_0\dot{\varphi}_1 = -\left(\frac{\partial\Phi_1}{\partial R}\right)_{R_0}, \label{eq:apd_a_R1} \\
& \ddot{\varphi}_1 + 2\Omega_0\frac{\dot{R}_1}{R_0} = -\frac{1}{R_0^2}\left(\frac{\partial\Phi_1}{\partial\varphi}\right)_{R_0}, \label{eq:apd_a_phi1}
\end{empheq}
\end{subequations}

To solve the above equation, we now adopt a specific logarithmic perturbing potential, i.e.,
\begin{equation}\label{eq:sp_pot}
    \Phi_1(R, \varphi) = \mathcal{A}(R) \exp \left[ i \left( m \varphi + m \frac{\ln(R/h_{R,1})}{\tan p} \right) \right],
\end{equation}
Let $m$ be the azimuthal wavenumber. The scale length of the potential perturbation is $h_{R,1}$, and the corresponding amplitude is $\mathcal{A}(R)$. The pitch angle of the spiral arms is characterized by $p$. When $\varphi_1\ll1$, we have $\varphi\approx\varphi_0(t)=\phi_0 - \Omega_p t$. 

Next, we integrate equation \eqref{eq:apd_a_phi1} to obtain $\dot{\varphi_1}$, which is then substituted into equation \eqref{eq:apd_a_R1} to yield the result:
\begin{align}
\ddot{R}_{1}+\kappa_{0}^{2}R_{1}=&-\left[\frac{2\Omega_{0}\mathcal{A}}{R_{0}(\Omega_{0}-\Omega_{p})}+\left(\frac{d\mathcal{A}}{dR}\right)_{R_{0}}\right]
\cos \chi+\frac{m\mathcal{A}}{R_{0}\tan p}\sin \chi,
\end{align}
where \( \chi = m(\phi_{0}-\Omega_{p}t) + m\frac{\ln(R_{0}/h_{R,1})}{\tan p} \). The epicycle frequency (or radial frequency) $\kappa_{0}^{2}=(R\frac{d\Omega^{2}}{dR} + 4\Omega^{2})_{R_0}$. The general solution to this equation, with $\Delta = \kappa_{0}^{2}-m^{2}(\Omega_{0}-\Omega_{b})^{2}$, is given by:  
\begin{align}
R_{1}(t)=&C_{0}\cos[\kappa_{0}t + \alpha]
-\left[\frac{2\Omega_{0}\mathcal{A}}{R_{0}(\Omega_{0}-\Omega_{p})}+\left(\frac{d\mathcal{A}}{dR}\right)_{R_{0}}\right]\frac{1}{\Delta}
\cos\chi+\frac{m\mathcal{A}}{R_{0}\tan p}\frac{1}{\Delta}\sin\chi,
\end{align}

In deriving the systematic response to the spiral potential, we retain only the particular solution (setting $C_0 = 0$, with only the second and third terms remaining) because the homogeneous part describes free epicyclic oscillations with random phases; these oscillations cancel out when averaging over a stellar population and do not contribute to the coherent velocity field associated with the density wave \citep[see also][]{schoenmakers1997measuring,siebert2012properties,eilers2020strength} The corresponding radial velocity is then given by:
\begin{align}\label{eq:ap_vr_cos}
V_{R,full}=&\left[\frac{2\Omega_{0}\mathcal{A}}{R_{0}(\Omega_{0}-\Omega_{p})}+\left(\frac{d\mathcal{A}}{dR}\right)_{R_{0}}\right]\frac{m(\Omega_{0}-\Omega_{p})}{\Delta}
\sin \chi+\frac{m\mathcal{A}}{R_{0}\tan p}\frac{m(\Omega_{0}-\Omega_{p})}{\Delta}\cos \chi,
\end{align}

\subsection{Surface density} \label{subsec:apd_sd1}
In a gravitational field, the surface density of a galaxy can be derived from the Poisson equation (see \citealt{binney2008galactic}, Section 2.3). Assuming that the 3D perturbed gravitational potential $\tilde{\Phi}_1(R,\phi,z)$ characterizes the surface density perturbation $\Sigma_1(R,\phi)$, the Poisson equation for a razor-thin disk (or a disk of finite thickness) can be expressed as:
\begin{align} \label{eq:poisson_3d}
    \nabla^2\tilde{\Phi}_1(R,\phi,z) &= 4\pi G\Sigma_1(R,\phi)\delta(z), \\
    \tilde{\Phi}_1(R,\phi,z) &= \Phi_1(R,\phi)\exp(-|z|/h_z),
\end{align}
Here, $h_z$ represents the scale height of the Galactic disk, which is assumed to be constant for all radius (i.e., $h_z = 1$ kpc). The operator $\nabla^2$ denotes the Laplacian in cylindrical coordinates, and $\delta(z)$ is the Dirac delta function. Integrating Eq.~\eqref{eq:poisson_3d} along the vertical direction from $z = -\infty$ to $z = +\infty$ yields:
\begin{equation}
    \frac{\partial^2\Phi_1(R,\phi)}{\partial R^2} + \frac{1}{R} \frac{\partial\Phi_1(R,\phi)}{\partial R} + \frac{1}{R^2} \frac{\partial^2\Phi_1(R,\phi)}{\partial \phi^2} = \frac{2\pi G\Sigma_1(R,\phi)}{h_z},
    \label{eq:poisson_2d}
\end{equation}

To solve Eq.~\eqref{eq:poisson_2d}, under the WKB approximation, we adopt the amplitude function $A(R)$ as given in \cite{eilers2020strength}:
\begin{equation}
    \mathcal{A}(R) = -\frac{2\pi G\Sigma_{\max}(R) R^2 \tan^2 p}{h_z m^2},
\end{equation}
where $\Sigma_{\max}(R)$ is the maximum surface density of the spiral arms, expressed as:
\begin{equation}
    \Sigma_{\max}(R) = \Sigma_{\max}(R_{\odot})\exp\left(-\frac{R - R_{\odot}}{h_{R,1}}\right),
\end{equation}
where $h_{R,1}$ is the  scale length of the perturbed disk. Based on the WKB approximation, we retain only the terms proportional to $\tan^{-2} p$, and the surface density perturbation can then be written as:

\begin{equation}\label{eq:density1}
    \Sigma_1(R,\phi) = \Sigma_{\max}(R) \cos \chi.
\end{equation}

\setcounter{figure}{0}
\renewcommand{\thefigure}{C\arabic{figure}}
\section{Parameter degeneracies} \label{sec:ap_blindsearch}

\begin{figure*}
    \centering
    \includegraphics[width=\textwidth]{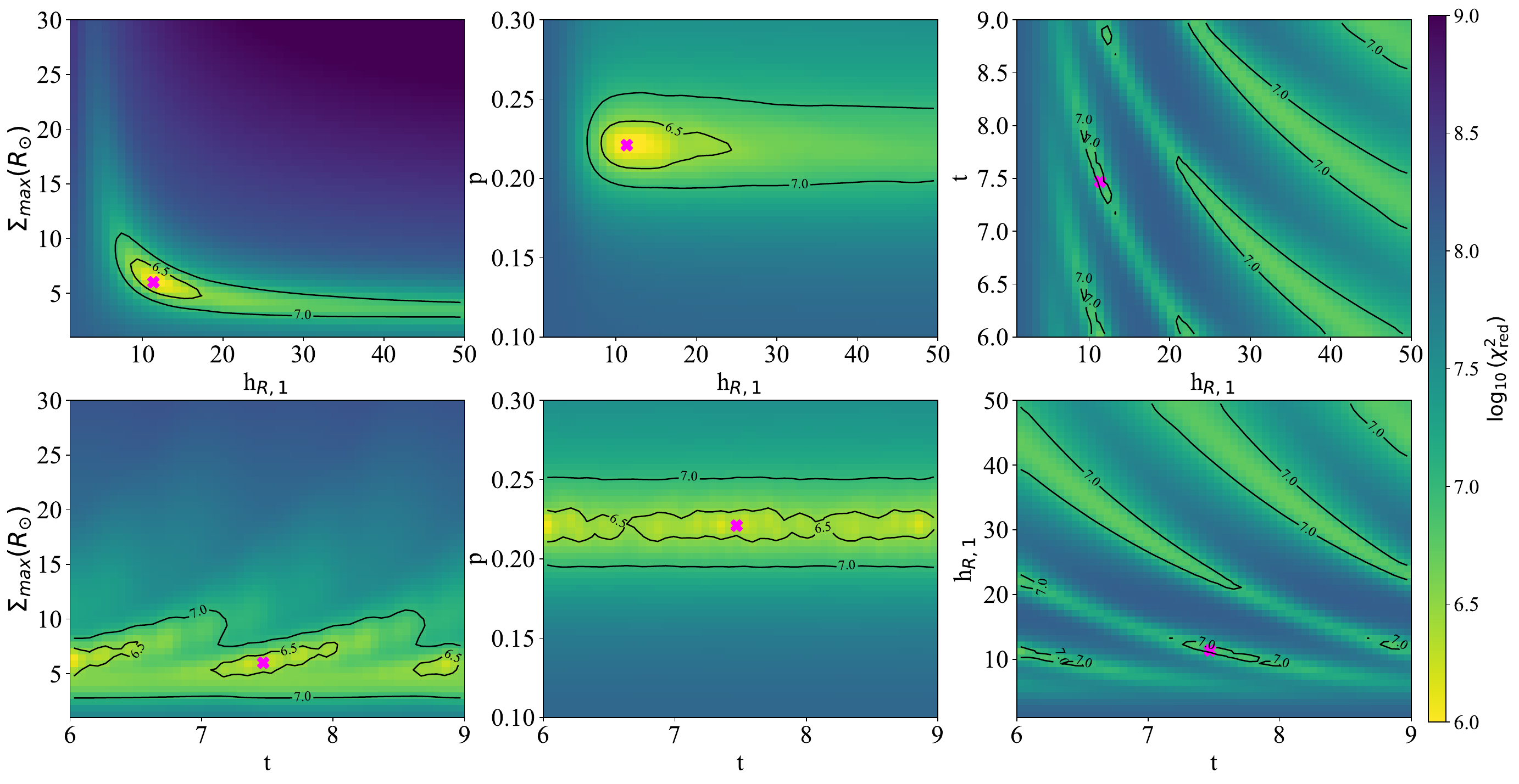}
    \caption{Two-dimensional maps of the logarithmic reduced chi-square $\log_{10}\chired$ for pairs of model parameters. The top panels show projections involving $\hr$, while the bottom panels show those involving $t$. The red crosses mark the true values adopted in the mock simulation.}
\label{fig:ap_blindsearch}
\end{figure*}

To visualize the parameter dependencies among $\fullparam$ we construct two-dimensional maps of the logarithmic reduced chi-square $\log_{10}\chired$ evaluated in the vicinity of the true simulation values (11.36,6,7.47,12.66)$\fullunit$ based on Eq.~(\ref{eq:vr_full}). The resulting pairwise projections are presented in Fig.~\ref{fig:ap_blindsearch}. The top row illustrates the dependence of $\hr$ on $\sigmamax$, $p$ and $t$, while the bottom row shows how $t$ correlates with 
$\sigmamax$, $p$ and $\hr$. We find that $\sigmamax$ and $p$ are well constrained by the model, whereas the joint distribution of $\hr$ and 
$t$ exhibits a clear periodic pattern and strong degeneracy, indicating that these two parameters cannot be independently determined within the current framework.

\setcounter{figure}{0}
\renewcommand{\thefigure}{D\arabic{figure}}
\section{Effects of scale length and local peak surface density} \label{sec:ap_hr_sfd}

\begin{figure*}
    \centering
    \includegraphics[width=\textwidth]{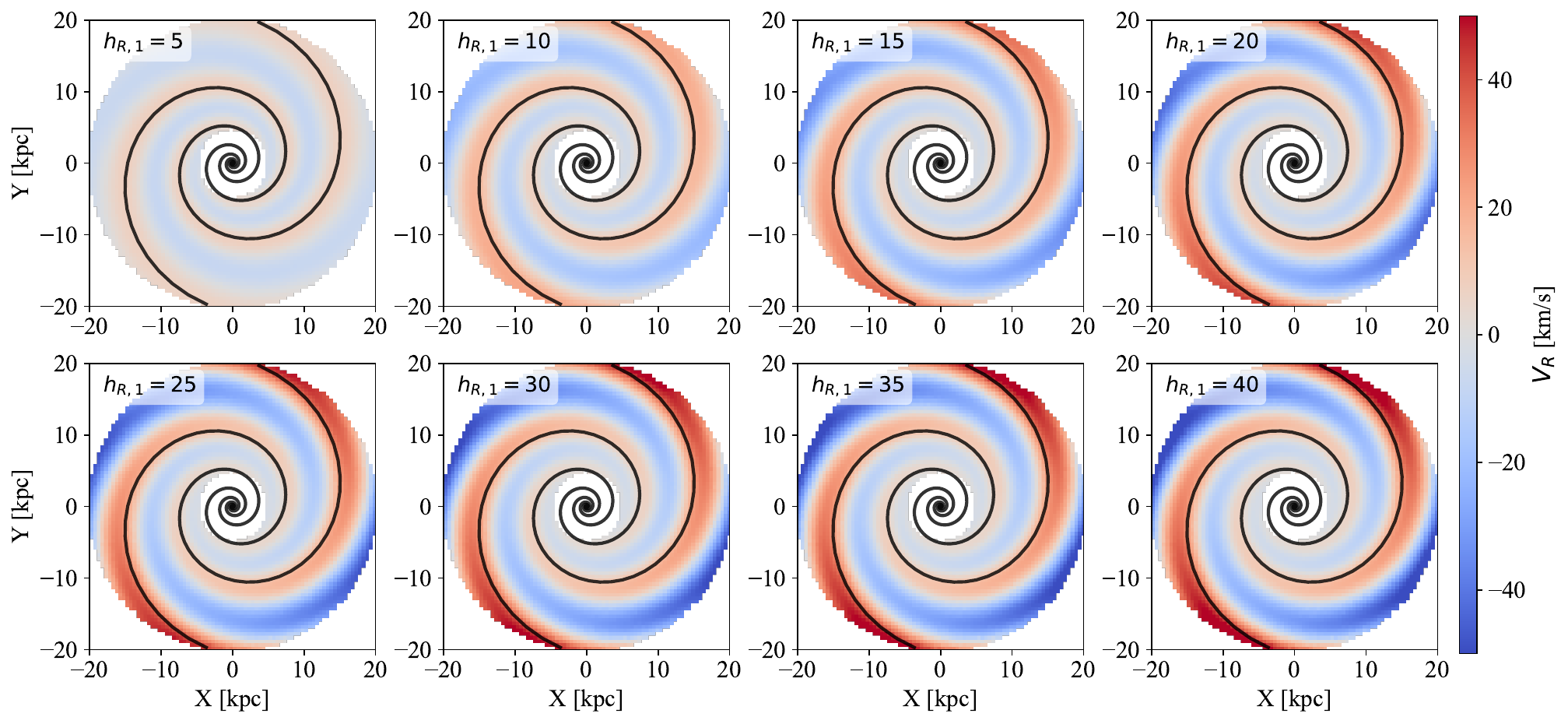}
    \caption{Results from the $V_R$ model with identical parameters but varying radial scale length from $\hr = 5$ to $40\kpc$ in steps of $5\kpc$. Black spiral curves denote the loci of the minimum gravitational potential.}
\label{fig:ap_hr1} 
\end{figure*}

\begin{figure*}
    \centering
    \includegraphics[width=\textwidth]{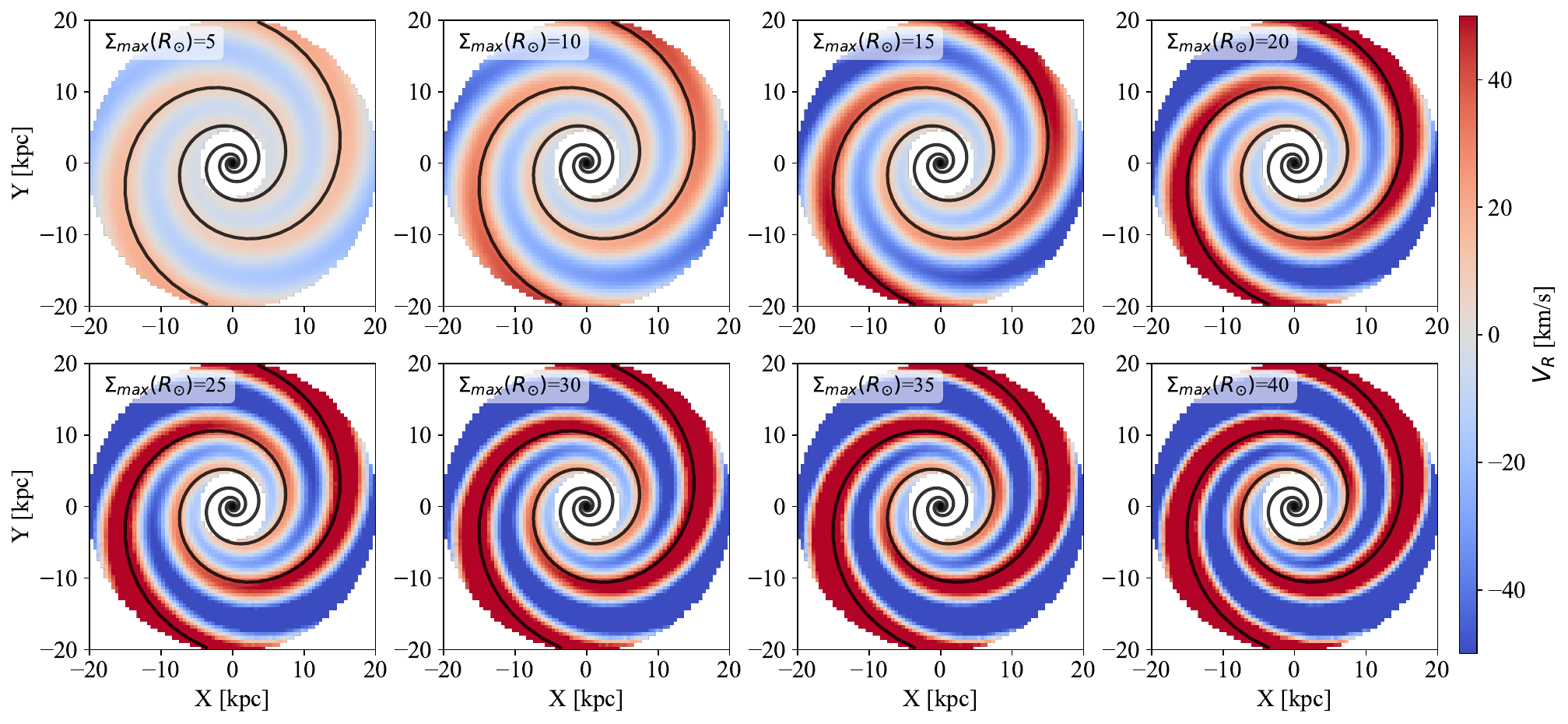}
    \caption{Same as Fig.~\ref{fig:ap_hr1}, but varying the local peak surface density $\sigmamax$.}
\label{fig:ap_sfd} 
\end{figure*}

Figures~\ref{fig:ap_hr1} and \ref{fig:ap_sfd} illustrate the effects of the radial scale length $\hr$ and the local peak surface density $\sigmamax$ on the $V_R$ maps in the corotating reference frame. Both $\hr$ and $\sigmamax$ primarily regulate the overall amplitude of the radial velocity signal. For otherwise identical parameter configurations, variations in $\sigmamax$ have a slightly stronger impact on the $V_R$ amplitude than changes in $\hr$. It is worth noting that, because the maps are constructed in the corotating frame, the spiral-arm loci remain fixed across different models. In an inertial reference frame, however, $\hr$ additionally affects the azimuthal phase of the disk response through the rotation term in Eq.~(\ref{eq:vr_full_sincos}), leading to a strong degeneracy with the spiral rotation time $t$.

\end{appendix}
\end{document}